\documentclass[amsmath,12pt,amssymb,preprint,nofootinbib,aps]{revtex4-2}
\usepackage{amsfonts} 
\usepackage{graphicx} 
\usepackage{epsfig}
\usepackage{multirow}
\usepackage{bm}
\usepackage{color}
\usepackage{amssymb}
\usepackage{dsfont}
\usepackage{hyperref}
\usepackage{cleveref}

\newcommand{\diracslash}[1]{#1\llap{/\kern2pt}}

\newcommand{\be}{\begin{equation}}
	\newcommand{\ee}{\end{equation}}
\newcommand{\bea}{\begin{eqnarray}}
	\newcommand{\eea}{\end{eqnarray}}
\newcommand{\ba}[1]{\begin{array}{#1}}
	\newcommand{\ea}{\end{array}}

\newcommand{\bt}{\begin{tabular}}
	\newcommand{\et}{\end{tabular}}

\newcommand{\beas}{\begin{eqnarray*}}
	\newcommand{\eeas}{\end{eqnarray*}}

\begin{document}
\title{Finite size effects on the transport coefficients of strongly
    interacting QCD matter}

\author{Dhananjay Singh}
\email{snaks16aug@gmail.com}
\affiliation{Department of Physics, Dr. B R Ambedkar National Institute of Technology Jalandhar, 
	Jalandhar -- 144008, Punjab, India}

\author{Arvind Kumar}
\email{kumara@nitj.ac.in}
\affiliation{Department of Physics, Dr. B R Ambedkar National Institute of Technology Jalandhar, 
	Jalandhar -- 144008, Punjab, India}

	\begin{abstract}
	The role of finite volume effects on the various transport coefficients of strongly interacting quark matter is analyzed in the Polyakov chiral SU(3) quark mean field model (PCQMF) at finite temperatures and chemical potentials incorporating fermionic vacuum term. Using a non-zero lower momentum cutoff and two different forms of the Polyakov loop potentials with quark back reaction, we study the following viscous properties: specific shear viscosity ($\eta/s$), normalized bulk viscosity ($\zeta_b/s$), and conductivity properties: electrical conductivity ($\sigma_{el}/T$), thermal conductivity ($\kappa/T^2$). Along with this, some essential thermodynamic quantities in the context of transport properties, such as the square of the speed of sound ($c_s^2$) and the specific heat ($c_v$) at a constant volume, are computed. Finite size effects are applied to the vacuum term and its influence on the effective quark masses, thermodynamic quantities, and transport coefficients is studied. The temperature dependence of the transport coefficients is obtained through the kinetic theory approach with the relaxation time approximation. The size of the system has been found to have significant effects on all transport coefficients. We find that all the transport coefficients increase as the size of the system is reduced. We have also studied the specific sound channel $(\eta+3\zeta_b/4)/s$ and the bulk-to-shear viscosity ratio $\zeta_b/\eta$. The effect of finite size is found to be more prominent in the transition region and vanishes at high $T$. The transition temperature $T_{\chi}$ is found to decrease as the system size (characterized by $R$) decreases. At finite chemical potentials, $T_{\chi}$ is shifted to lower values compared to the case of the vanishing chemical potential.

	\end{abstract}
	
	\maketitle
\newpage	
	\section{\label{intro}Introduction}
The investigation of transport coefficients in strongly interacting matter at finite temperatures and baryonic densities is a topic of increasing interest, with applications in diverse scenarios. 
This inquiry is of particular relevance in the context of relativistic heavy-ion collisions where elevated temperatures and lower densities are reached \cite{brat}. Notably, experiments at the Relativistic Heavy Ion Collider (RHIC) at BNL and the Large Hadron Collider (LHC) at CERN have generated deconfined states of quarks and gluons, known as quark-gluon plasma (QGP) \cite{roma,luzum,roy}. In contrast, at low temperatures and high densities, transport coefficients may play a significant role in explaining many astrophysical phenomena in compact stars \cite{buballa,baym}. The experimental facilities at the Nuclotron-based Ion Collider Facility (NICA) at JINR, Russia \cite{jinr} and the Facility for Antiproton and Ion Research (FAIR) at GSI, Germany \cite{fair} are expected to probe matter at such densities and temperatures. Transport coefficients such as shear viscosity $\eta$, bulk viscosity $\zeta_b$, electrical conductivity $\sigma_{el}$, and thermal conductivity $\kappa$ offer invaluable insight into its collective flow, compressibility, and response to electromagnetic fields \cite{heinz,patra}. The QGP exhibits a specific shear viscosity (shear viscosity to entropy density ratio, $\eta/s$) that nearly approaches the lower limit ($\eta/s = 1/4\pi$), famously known as the Kovtun-Son-Starinets (KSS) bound \cite{kss}, an attribute of strongly coupled media. However, due to the property of asymptotic freedom, perturbative calculations in the high-temperature domain yield a value of $\eta/s$ that surpasses the lower bound by a factor of 10-20 \cite{arnold,arnold03}, seemingly indicative of a weakly interacting medium. This discrepancy between experimental and theoretical values has spurred the utilization of alternative methodologies to probe the viscous properties of hot QCD matter arising from heavy-ion collisions.

\par The transport coefficient, the bulk viscosity $\zeta_b$ provides the information on the resistance to fluid volume or density changes. Some analytical calculations grounded in effective models have provided estimates for $\zeta_b$ in strongly interacting matter \cite{noro,sasaki2009,chandra,song}. Similarly, the electrical conductivity $\sigma_{el}$ has garnered substantial interest regarding heavy-ion collisions. This parameter governs the flow of charge carriers and is relevant in explaining the low-mass dimuon enhancement \cite{rapp,mohanty}, closely related to restoring chiral symmetry. Also, the transport coefficient, thermal conductivity $\kappa$, is of significance in high-density regimes where thermal conduction in the medium is influential. The behaviour of thermal conductivity $\kappa$ has been explored with various models in the literature \cite{marty,deb,fernandez,twasaki,mitra2015,mitra20152,shov,braby}. 
\par In recent times, a range of effective QCD \cite{sasaki2010,marty,ghosh2015,ghosh2016,deb,tawfik} and hadronic models \cite{itakura,fernandez,lang,mitra,ghosh2014,ghosh20142,ghoshb,kadam} have been used to gain deeper insight into various transport coefficients. Additionally, approaches such as transport simulations \cite{demir,pal,plumari}, Lattice QCD  \cite{meyer,astra}, and functional renormalisation group (FRG) \cite{christiansen} have contributed significantly. 
In Ref. \cite{ghosh2019}, the Nambu-Jona-Lasinio (NJL) model is used to study the role of elastic and inelastic scatterings on $\eta,\kappa,\sigma_{el}$ at vanishing as well as finite chemical potential and found that inelastic scattering exists only above the Mott temperature where the medium behaves like a perfect fluid. In Ref. \cite{sasaki2009}, authors studied the behaviour of $\zeta_b$ near the phase transition using quasiparticle approaches like the Ginzburg-Landau model and the scaling theory and showed that $\zeta_b$ is not sensitive to the chiral phase transition. Polyakov linear sigma model (PLSM) model was used to study the temperature dependence of the transport coefficients in Ref. \cite{tawfik}. In Ref. \cite{ozven}, the parton-hadron-string-dynamics (PHSD) approach is used to study $\eta$ and $\zeta_b$ of partonic and hadronic matter and found that $\eta/s$ shows a minimum while $\zeta_b/s$ shows a significant rise near the critical temperature $T_c$. Authors in Ref. \cite{islam} used different versions of the NJL model, like the PNJL model and the EPNJL model, to study the transport coefficients and found that they contain the dynamics of both the chiral and deconfinement phase transitions. In Ref. \cite{patra}, the effect of the strong magnetic field on the $\sigma_{el}$ and $\kappa$ is studied using the quasiparticle approach and found that it causes drastic enhancement in them. The impact of non-extensivity on the transport coefficients was analyzed in Ref. \cite{rath} using the kinetic theory approach. The authors found that $\sigma_{el}$ and $\kappa$ increase with increasing non-extensivity.  
A comprehensive synthesis of these coefficients' temperature dependencies reveals a decrease in $\eta(T)$, $\sigma_{el}(T)$, and $\kappa(T)$ within the hadron phase, contrasting with an increase in these quantities as a function of temperature in the quark phase\cite{sasaki2010,ghosh2015,deb,fernandez}. An opposite trend is observed for $\zeta_b(T)$ \cite{paech, kharzeev, karsch2008}. Hence, near the transition temperature, a minimum is expected in the value of $\eta(T)$, $\sigma_{el}(T)$, and $\kappa(T)$ \cite{gavin,prakash,davesne,chakraborty,cassing,fernandez2006,ghosh17}. In contrast, a maximum is expected in $\zeta_b(T)$ \cite{chakraborty,sasaki2010,song,saha17}. These collective findings underscore the crucial role of transport coefficients in characterising the properties of QCD matter as a strongly interacting fluid.

The QGP produced in heavy-ion collisions exists within a confined volume, potentially influencing its dynamical properties and, consequently, its transport coefficients. Surprisingly, the close value of QGP's $\eta/s$ to the KSS bound suggests that the matter is susceptible to quantum fluctuations \cite{saha2018}. Therefore, finite volume effects may play a significant role in studying transport coefficients. To comprehend the impact of finite size effects, it is essential to recognise the intricate dynamics of the fireball generated immediately after a heavy-ion collision. This dynamic process hinges on various factors, such as the size of the colliding nuclei, the energy of their centre of mass, and the impact parameter. The relativistic nuclear collision has been reported to create a hadronic fireball with a radius of $2 $ fm \cite{koch}. The QGP matter produced in LHC and RHIC has been estimated to have a size range of $2-10 $ fm \cite{braun,palhares}. Studies of finite size effects on thermodynamics and phase transitions of QCD matter have already demonstrated their considerable significance \cite{nisha,fisher,gasser,gopie,kiriyama,yasui,luecker,abreu,bhatt2013,karsch,bhatt2016}.
This motivates us to investigate how the intricate dynamics of the fireball, shaped by factors like colliding nuclei and impact parameters, influence the transport coefficients of the system, which serve as invaluable tools in characterising the nature of phase transition \cite{sasaki2009}.

\par The finite size effects for strongly interacting medium can be incorporated in the effective models by two methods: by choosing the appropriate boundary conditions \cite{abreu2017,davoudi,gasser1988,carrizal,zhang2020} or by implementing a lower momentum infrared cutoff \cite{bhatt2013,karsch,samanta,sarkar}. Finite size effects in the finite-temperature field theory are introduced by compactifying the four-momentum variables $(k_0,\vec{k})$. For the time component, this is done through the imaginary time formalism in which one introduces an imaginary time $\tau \in [0,i\beta]$, where $\beta=T^{-1}$ \cite{abreu}. This corresponds to the generalized Matsubara prescription, which makes the energy component discretized as $k_0 \rightarrow \omega_{n_{0}} = \frac{2\pi}{\beta}(n_0+\frac{1}{2})$, with $n_0=0,\pm 1,\pm 2,..$ satisfying the anti-periodic boundary conditions (APBC) for the fermion field. For the spatial coordinates, we restrict ourselves to finite size or length $R, i.e., \vec{k}\in[0,R]$. This corresponds to discretizing the momentum component as $\vec{k} \rightarrow \omega_n = \frac{2\pi}{R}(n+c)$, with $n=0,\pm 1,\pm 2,..$ \cite{abreu}. Choosing the boundary conditions of the fermion fields in the spatial direction means choosing the value of $c$ such that $c=0,1/2$ corresponds to APBC and PBC, respectively. Finite size effects with APBC and PBC boundary conditions for the fermionic field have been incorporated into various models \cite{abreu2017,davoudi,gasser1988,braun2005}. Finite size effects can also be implemented using the stationary wave condition (SWC), which makes the quark wave function zero on the boundary \cite{wang2018}.
Authors in Ref. \cite{karsch} used a one dimensional gas of non interacting bosons and showed that the discrete sum in the momentum can be made continuous by implementing a lower momentum cut off.
In the Hadron Resonance Gas (HRG) model, the finite size effects with lower momentum cutoff have been used to study the transport properties of hadronic matter \cite{samanta,sarkar}. The transport coefficients of quark matter were analysed in a finite volume Polyakov Nambu Jona Lasinio (PNJL) model, revealing a notable effect as the system size varied \cite{saha2018}. In addition to the momentum discretization, the finite size effects should also contain surface and curvature effects \cite{shao}. In the NJL model, the finite size effects have been incorporated by considering spherical and cubic regions \cite{carrizal}. Finite volume effects have been taken into consideration by the spherical boundary condition in the MIT bag model \cite{zhang2020}. In the PNJL model, finite size effects have been studied by employing multiple reflection expansion (MRE) formalism where the density of states is modified by considering a spherical rather than a cubic volume \cite{grunfeld,kiriyama}.

\par We intend to investigate the transport coefficients within the finite volume Polyakov chiral SU(3) quark mean field (PCQMF) model, which effectively describes QCD matter's thermodynamics and phase transition across a range of values for chemical potentials and temperatures \cite{manisha}. To introduce finite volume effects in the PCQMF model, we have implemented a lower momentum cutoff, denoted as $k_{min} = \frac{\pi}{R} = \lambda$ (where $R$ is the length of the cubic volume) in the integrals of the equations of motion and transport coefficients. We have neglected the surface and curvature effects for simplicity. In addition, we have examined the influence of two different forms of the Polyakov loop potentials in the presence of the quark back reaction. The presence of dynamical quarks alters gluons' dynamics, replacing the usual Polyakov loop potential with the QCD glue potential. Incorporating the quark back reaction in the Polyakov loop significantly improves their agreement with lattice data \cite{lisa, suraj}. The effect of the quark-improved Polyakov loop potential on the thermodynamics and transport properties of the quark matter is studied. Furthermore, we have incorporated the fermionic vacuum term into the framework of the PCQMF model. A detailed examination of how the vacuum term is affected by finite size effects has been conducted. We compute the transport coefficients across various temperatures and densities using the kinetic theory approach and the relaxation time approximation. We incorporate the quark back reaction by replacing the pure gauge temperature $T_{YM}$ with the glue potential temperature $T_{glue}$. Finite size effects alter the in-medium dispersion relations, subsequently modifying the transport coefficients.

\par The article is organised as follows: In Sec. \ref{pcqmf}, we briefly introduce the Polyakov extended chiral SU(3) quark mean field model. The Polyakov loop potentials used in the present work are introduced in Sec. \ref{polyakov}. The implementation of finite size effects in the PCQMF model are discussed in Sec. \ref{finitevolume}. Mathematical expressions of the transport coefficients: shear viscosity ($\eta$), bulk viscosity ($\zeta_b$), electrical conductivity ($\sigma_{el}$), and thermal conductivity ($\kappa$) are discussed in Sec. \ref{transport}. In Sec. \ref{results}, the impact of finite size and various chemical potentials on the transport coefficients are presented. Finally, in Sec. \ref{summary}, we summarise the results of the present work.
	
	\section{\label{method} Methodology }
\subsection{\label{pcqmf}\textbf{Polyakov chiral SU(3) quark mean field model}}
We employ the PCQMF model in finite volume to investigate the temperature dependence of various transport coefficients of the QGP matter. This model incorporates quark-meson and meson-meson interactions and is based on QCD's broken scale invariance property \cite{papa1999,wang2001,wang2002}.
The PCQMF model is an extension of the chiral SU(3) quark mean field (CQMF) model in which the Polyakov loop is introduced to study the deconfinement transition. The PCQMF model has been first applied to study the thermodynamics and phase transition of quark matter \cite{manisha}. The fluctuations of conserved charges have been studied using this model in Ref. \cite{nisha22}. The effect of finite size on the thermodynamic properties has been studied using the PCQMF model in Ref. \cite{nisha}. This model has been recently used to study the impact of non-extensivity on the chiral and deconfinement phase transition of the QCD matter \cite{dj}. At finite temperatures and chemical potential, the model describes the interactions between quarks by exchanging meson fields (both scalar and vector). Specifically, the scalar meson fields $\sigma$, $\zeta$, and $\delta$ which are responsible for the attractive part of the interactions, the vector meson fields $\omega$, $\rho$ and $\phi$ accounting for the repulsive interactions, and the Polyakov fields $\Phi$ and $\bar{\Phi}$ for investigating the characteristics of deconfinement phase transitions are considered in the present study. Furthermore, the model introduces a scalar dilaton field, $\chi$, also called the glueball field, to incorporate the phenomenon of broken scale invariance \cite{schechter,gomm}. The effective Lagrangian of the PCQMF model for three flavoured quark matter consists of two parts \cite{manisha}
\begin{equation}
	{\cal L}_{{\rm PCQMF}} \, = \, {\cal L}_{\rm chiral} \, -{\cal U}(\Phi(\vec{x}),\bar{\Phi}(\vec{x}),T), \label{totallag}
\end{equation}
where the chiral part is represented by

\begin{equation}
	\label{lag_chiral}
	{\cal L}_{{\rm chiral}} \, = \, {\cal L}_{q0} \, + \, {\cal L}_{qm}	\, + \,	{\cal L}_{\Sigma\Sigma} +{\cal L}_{VV} +{\cal L}_{SB}\, + \, {\cal L}_{\Delta m} \,
\end{equation}
consisting of the fermionic part, ${\cal L}_{q0} = \bar \psi \, i\gamma^\mu \partial_\mu \psi$, with $\psi=\left(u,d,s\right)$.
${\cal L}_{qm}$ represents the interactions between quark and meson and is given by
\begin{eqnarray}
	{\cal L}_{qm}  =  g_s\left(\bar{\psi}_LM \psi_R+\bar{\psi}_RM^+\psi_L\right)- g_v\left(\bar{\psi}_L\gamma^\mu l_\mu \psi_L+\bar{\psi}_R\gamma^\mu r_\mu \psi_R\right), 
\end{eqnarray}
where $g_v$$(g_s$) are vector (scalar) coupling constants. The spin-0 scalar ($\Sigma$) and pseudoscalar ($\Pi$) mesons are expressed as \cite{papa1999}
\begin{equation}
	M(M^{\dagger}) = \Sigma \pm i\Pi = \frac{1}{\sqrt{2}}\sum_{a=0}^{8}(\sigma^a\pm i\pi^a)\lambda^a,
\end{equation}
where $\sigma^a (\pi^a)$ represent the scalar (pseudoscalar) meson nonets and $\lambda^a$ are the Gell-Mann matrices with $\lambda^0=\sqrt{\frac{2}{3}}\mathds{1}$. Similarly, spin-1 vector ($V_{\mu}$) and pseudovector ($A_{\mu}$) mesons are introduced through \cite{papa1999}  
\begin{equation}
	l_{\mu}(r_{\mu}) = \frac{1}{2}(V_{\mu}\pm A_{\mu}) = \frac{1}{2\sqrt{2}}\sum_{a=0}^{8}(v^a_{\mu}\pm a^a_{\mu})\lambda^a,
\end{equation}
with $v^a_{\mu}$ ($a^a_{\mu}$) being vector (pseudovector) meson nonets \cite{papa1999}. The interaction Lagrangian ${\cal L}_M$ is described by,
\begin{equation}
	\label{lag_meson}
	{\cal L}_M = {\cal L}_{\Sigma\Sigma} +{\cal L}_{VV} +{\cal L}_{SB},
\end{equation}
with the scalar meson self-interaction term ${\cal L}_{\Sigma\Sigma}$ in the mean-field approximation is expressed as			
\begin{eqnarray}
	{\cal L}_{\Sigma\Sigma} =& -\frac{1}{2} \, k_0\chi^2
	\left(\sigma^2+\zeta^2+\delta^2\right)+k_1 \left(\sigma^2+\zeta^2+\delta^2\right)^2\nonumber \\
	&+k_2\left(\frac{\sigma^4}{2} +\frac{\delta^4}{2}+3\sigma^2\delta^2+\zeta^4\right) +k_3\chi\left(\sigma^2-\delta^2\right)\zeta 
	-k_4\chi^4\nonumber \\
	&-\frac14\chi^4 {\rm ln}\frac{\chi^4}{\chi_0^4} +
	\frac{d}
	3\chi^4 {\rm ln}\left(\left(\frac{\left(\sigma^2-\delta^2\right)\zeta}{\sigma_0^2\zeta_0}\right)\left(\frac{\chi^3}{\chi_0^3}\right)\right).
	\label{scalar0}
\end{eqnarray}			
The PCQMF model is based on the broken scale invariance of QCD, which is introduced through a scale-breaking term in the Lagrangian density \cite{schechter,gomm,heide}. The last two logarithmic terms in the above equation introduce this scale-breaking effect
\cite{wang2003}. The vacuum value of $\sigma$ and $\zeta$ fields is $\sigma_0 = - f_\pi = -93$ MeV and $\zeta_0  = \frac{1}{\sqrt{2}} ( f_\pi - 2 f_K) = -95.47$ MeV, respectively. Here, $f_\pi$ and $f_K$ are the pion and kaon decay constants, respectively. The term ${\cal L}_{VV}$ in Eq. (\ref{lag_meson}) describes the vector meson self-interaction and is given by
\begin{equation}
	{\cal L}_{VV}=\frac{1}{2} \, \frac{\chi^2}{\chi_0^2} \left(
	m_\omega^2\omega^2+m_\rho^2\rho^2+m_\phi^2\phi^2\right)+g_4\left(\omega^4+6\omega^2\rho^2+\rho^4+2\phi^4\right), \label{vector}
\end{equation}
with $m_{\omega}, m_{\rho}, m_{\phi}$ being the vector meson masses which can be written as \cite{wang2003}
\begin{eqnarray}
	m^2_\omega = m_\rho^2 =
	\frac{m_v^2}{1 - \frac{1}{2} \mu \sigma^2}\, ,
	\hspace*{.5cm} {\rm and} \hspace*{.5cm}
	m^2_\phi = \frac{m_v^2}{1 - \mu \zeta^2}\, .
\end{eqnarray} 
The vacuum mass of the vector meson $m_v$ = 673.6 MeV and density parameter $\mu$ = 2.34 fm$^2$ are fixed to produce ${m_\phi}$ = 1020 MeV and ${m_\omega}$ = 783 MeV. The term, ${\cal L}_{SB}$ in Eq.(\ref{lag_chiral}) represents the spontaneous symmetry-breaking term responsible for giving masses to pseudoscalar mesons and is written as \cite{wang2002,wang2004,papa1999}
\begin{equation} 
	{\cal L}_{SB}=-\frac{\chi^2}{\chi_0^2}\left(h_x\sigma + 
	h_y\zeta\right),
	\label{esb_ldensity}
\end{equation} 
where $h_x= m^2_{\pi}f_{\pi}$ and $h_y = (\sqrt{2}m_K^2f_K-\frac{1}{\sqrt{2}}m_\pi^2f_\pi)$. The term ${\cal L}_{\Delta m}$ in Eq. (\ref{lag_chiral}) is an additional mass term which helps to generate the exact mass of $s$ quark, is represented by ${\cal L}_{\Delta m} = - \Delta m_s \bar \psi S \psi$,
where $\Delta m_s = 29$ MeV and $S \, = \, \frac{1}{3} \, \left(I - \lambda_8\sqrt{3}\right) = {\rm diag}(0,0,1)$.

\subsubsection{\label{polyakov} Polyakov loop potentials}
\par The Polyakov loop potential mimics the quark-gluon interactions and serves as a tool for studying deconfinement phase transition by introducing the Polyakov field $\Phi$ and its conjugate $\bar{\Phi}$ \cite{ratti,fukushima}. The functional form of the effective Polyakov loop potential is not unique and depends on the centre symmetry of the pure gauge sector. The various parameters associated with distinct Polyakov loop potentials can be determined by analysing pure gauge lattice data when the chemical potential is set to zero \cite{pisarski}. For the present study, we have used two forms of Polyakov loop potentials: the polynomial form Polyakov loop potential ${\cal U_P}(\Phi, \bar{\Phi})$ and the logarithm form Polyakov loop potential ${\cal U_L}(\Phi, \bar{\Phi})$. The polynomial parameterised effective Polyakov loop potential is defined as \cite{ratti,scavenius}
\begin{equation}
	\frac{{\cal U_P}(\Phi, \bar{\Phi})}{T^4} = -\frac{b_2(T)}{2}\bar{\Phi}\Phi - \frac{b_3}{6}(\Phi^3+\bar{\Phi}^3) + \frac{b_4}{4}(\bar{\Phi}\Phi)^2,
	\label{polypoly}
\end{equation}
with a temperature-dependent coefficient
\begin{equation}
	b_2(T) = a_0 + a_1\bigg(\frac{T_0}{T}\bigg) + a_2\bigg(\frac{T_0}{T}\bigg)^2 + a_3\bigg(\frac{T_0}{T}\bigg)^3.
\end{equation}
The corresponding parameters which are fitted using lattice data are $a_0 = 6.75,$ $a_1 = -1.95,$ $a_2 = 2.625,$ $a_3 = -7.44,$ $b_3 = 0.75,$ $b_4 = 7.5$ \cite{suraj}. 
\par The polynomial form of the Polyakov loop potential is improved by introducing the SU(3) Haar measure \cite{ghosh2008}, which introduces a Jacobian determinant. The logarithm of this determinant is included as an effective potential in the action's generating functional \cite{fukushima}. The resulting logarithmic form of effective Polyakov loop potential considered in the present work is written as \cite{fukushima,costa,roessner}
\begin{eqnarray}
	\frac{{\cal U_L}(\Phi,\bar{\Phi})}{T^4}&=&-\frac{a(T)}{2}\bar{\Phi}\Phi+b(T)\mathrm{ln}\big[1-6\bar{\Phi}\Phi+4(\bar{\Phi}^3+\Phi^3)-3(\bar{\Phi}\Phi)^2\big],
	\label{log}
\end{eqnarray}
with the temperature-dependent coefficients:
\begin{equation}\label{T}
	a(T)=a_0+a_1\bigg(\frac{T_0}{T}\bigg)+a_2\bigg(\frac{T_0}{T}\bigg)^2,\ \  b(T)=b_3\bigg(\frac{T_0}{T}\bigg)^3.
\end{equation}
The corresponding parameters are: $a_0 = 1.81$, $a_1 = -2.47$, $a_2 = 15.2$ and $b_3 = -1.75$ \cite{manisha,nisha22}. Unlike ${\cal U_P}(\Phi, \bar{\Phi})$, the logarithmic form ${\cal U_L}(\Phi, \bar{\Phi})$ bounds the expectation value of $\Phi,\bar{\Phi}$ from below at high temperatures \cite{fukugita,zhang}. This is because the logarithmic term in Eq. (\ref{log}) diverges as $\Phi,\bar{\Phi}\rightarrow 1$ and therefore $\Phi,\bar{\Phi}$ remains smaller than 1 and approaches this value asymptotically as $T\rightarrow\infty$.

The parameter $T_0=270$ MeV is the critical temperature for the phase transition of deconfinement in the pure gauge sector ($\mu = 0$)  \cite{fukugita,schaefer}. However, in the presence of fermionic fields, $T_0$ becomes dependent on $N_f$ and is adjusted to lower values to produce better lattice results \cite{schaefer,herbst}.
The behaviour of the gluons responsible for creating the Polyakov loop potential changes due to the inclusion of dynamical quarks \cite{lisa}. Including quark back reaction results in replacing the Polyakov loop potential with the QCD glue potential. In Ref. \cite{lisa}, authors include the quark back reaction within the Polyakov quark meson (PQM) model and found significant differences between the results of the pure gauge Polyakov loop potential ${\cal U}_{YM}$ and the glue Polyakov loop potential ${\cal U}_{glue}$. Ref. \cite{suraj} used the renormalized 2 + 1 flavour quark meson model (RQM) to study the chiral and deconfinement phase transition. They found that the inclusion of the quark back reaction results in shifting the CEP to lower values of temperatures while the shift in the chemical potential is negligible. The temperature of the pure gauge system $T_{YM}$ is related to the temperature of the glue potential $T_{glue}$ as \cite{lisa} 
\begin{equation}\label{Tglue}
	\frac{T_{YM} - T_{0}^{YM}}{T_{0}^{YM}} = 0.57 \frac{T_{glue} - T_{0}^{glue}}{T_{0}^{glue}}.
\end{equation}
Denoting ${\cal U}_{YM}$ as the Polyakov loop potential used so far, the improved glue Polyakov loop potential, ${\cal U}_{glue}$ can be constructed as 
\begin{equation}\label{Uglue}
	\frac{{\cal U}_{glue}(\Phi,\bar{\Phi},T_{glue})}{T_{glue}} = \frac{{\cal U}_{YM}(\Phi,\bar{\Phi},T_{YM})}{T_{YM}}.
\end{equation}
For the replacement procedure, we have used $T \rightarrow T_{0}^{YM}\left(1 + 0.57\left(\frac{T_{glue}}{T_{0}^{glue}-1}\right)\right)$ in the RHS of the Polyakov loop potentials
with $T_0$ being $T_{0}^{YM}$ \cite{suraj}.
We have taken $T_{0}^{YM} = T_{0}^{glue} = 200$ MeV in our calculations.


\subsubsection{Finite size effects}
\label{finitevolume}
\par As discussed in Sec. \ref{intro}, we have used the lower momentum infrared cutoff to introduce the finite size effects within the PCQMF model.
The infinite sum over discrete momentum values is approximated by integration over continuous momentum variation while retaining the infrared cutoff. 
In the present work, we have introduced the fermion vacuum term in the PCQMF model, which is found to have a critical role in PQM, NJL, and PNJL models \cite{saha2018,tiwari2012,skokov}. 
Owing to these changes in the mean-field approximation, the thermodynamic potential density of the PCQMF model is given by 
\begin{equation}
	\hspace*{-.4cm} 
	\Omega_{\rm{PCQMF}}= \Omega_{vac} + \mathcal{U}(\Phi,\bar{\Phi},T) - {\cal L}_M- {\cal V}_{vac} + \sum_{i=u,d,s}\frac{-\gamma_i k_BT}{(2\pi)^3}\int_{\lambda}^{\infty} 
	d^3k\left\{ {\rm ln} 
	F^{-}+
	{\rm ln} F^{+}\right\}.
	\label{tpdpcqmf}
\end{equation}
In the above equation, the term $\Omega_{vac}$ is the fermion vacuum contribution term expressed as
\begin{eqnarray}
    \Omega_{vac} = -2N_c \sum_{i=u,d,s}\int_{\lambda}^{\infty}\frac{d^3k}{(2\pi)^3}E_i^*(k).
\end{eqnarray}
The above integral is regularized by employing dimensional regularization within the finite volume to obtain \cite{kovacs}
\begin{eqnarray}
    \Omega_{vac} = -\frac{N_c}{8\pi^2}\sum_{i=u,d,s}\left[m_i^{*4} {\rm ln} \left(\frac{\lambda+\sqrt{\lambda^2+m_i^{*2}}}{\Lambda_0}\right)-\lambda\sqrt{\lambda^2+m_i^{*2}}(2\lambda^2+m_i^{*2})\right].
    \label{omegavac}
\end{eqnarray}
Here, $\Lambda_0$ is the regularization scale parameter. It is important to note that if the lower momentum cutoff $\lambda$ is removed, the above equation reduces to 
\begin{eqnarray}
    \Omega_{vac}\big{|}_{\lambda\rightarrow0} = -\frac{N_c}{(8\pi^2)}\sum_{i=u,d,s}m_i^{*2}{\rm ln}\left(\frac{m_i}{\Lambda_0}\right),
\end{eqnarray}
which is the usual vacuum term obtained after dimensional regularization within infinite volume \cite{chatterjee2012}. In Ref. \cite{kovacs}, authors used the extended PQM model and observed that considering the finite size in the vacuum term leads to a decrease in the effective masses of quarks with decreasing size of the system at lower temperatures. 
They pointed out that what's crucial is not simply the existence of the vacuum contribution but rather how it is treated, i.e., its finite or infinite size.
    
The term $\mathcal{U}(\Phi,\bar{\Phi},T)$ in Eq. (\ref{tpdpcqmf}) represents the Polyakov loop potential discussed in Sec. \ref{polyakov}. The meson interaction term $\cal{L}_M$ is given by Eq. (\ref{lag_meson}). The subtraction of the term ${\cal V}_{vac}$ in Eq. (\ref{tpdpcqmf}) is performed to achieve zero vacuum energy. In the last term of Eq. (\ref{tpdpcqmf}) $\gamma_i$ = 2 is the spin degeneracy factor, and
\begin{eqnarray}
	F^{-}=&1+e^{-3E^-}+3\Phi e^{-E^-}+3\bar{\Phi}e^{-2E^-}, \\
	F^{+}=&1+e^{-3E^+}+
	3\bar{\Phi} e^{-E^+}
	+3\Phi e^{-2E^+},
\end{eqnarray} 
with $E^+ = (E_i^*(k)+{\mu_i}^{*})/k_BT$ and $E^- = (E_i^*(k)-{\mu_i}^{*})/k_BT$. Here, $E_i^*(k)=\sqrt{m_i^{*2}+k^2}$ represents the effective energy of a single quasiparticle. 

  The chemical potential of quarks within the medium ${\mu_i}^{*}$ is related to the chemical potential $\mu_i$ in free space through the relation  
\begin{equation}
	{\mu_i}^{*}=\mu_i-g_\omega^i\omega-g_\phi^i\phi-g_\rho^i\rho,
	\label{mueff}
\end{equation} 
where $g^i_{\omega}$, $g^i_{\phi}$, and $g^i_{\rho}$ represent the coupling coefficients between various quarks and the vector meson fields. The effective quark mass, ${m_i}^{*}$, is given by   
\begin{equation}
	{m_i}^{*} = -g_{\sigma}^i\sigma - g_{\zeta}^i\zeta - g_{\delta}^i\delta + \Delta m_i,
	\label{mbeff}
\end{equation}
with $g_{\sigma}^i$, $g_{\zeta}^i$, and $g_{\delta}^i$ representing the coupling constants that quantify the interaction strength between scalar fields and various quarks. At zero temperature and baryon chemical potential, quark masses are obtained by  $m_u = m_d = -\frac{g_s}{\sqrt{2}}\sigma_0$ and $m_s = -g_s\zeta_0+\Delta m_s$.
The parameters $g_{\sigma}^i$ and $g_{\zeta}^i$ are determined by adjusting them to $m_u= m_d$ = 313 MeV and $m_s = $ = 490 MeV \cite{wang2003}. 
The equations of motion of the fields are obtained after minimizing $\Omega$ in Eq. (\ref{tpdpcqmf}) with respect to $\sigma$, $\zeta$, $\delta$, $\chi$, $\omega$, $\rho$, $\phi$, $\Phi$, and $\bar{\Phi}$ as, 

\begin{equation}
	\label{minimize}
	\frac{\partial\Omega}
	{\partial\sigma}=\frac{\partial\Omega}{\partial\zeta}=\frac{\partial\Omega}{\partial\delta}=\frac{\partial\Omega}{\partial\chi}=\frac{\partial\Omega}{\partial\omega}=\frac{\partial\Omega}{\partial\rho}=\frac{\partial\Omega}{\partial\phi}=\frac{\partial\Omega}{\partial\Phi}=\frac{\partial\Omega}{\partial\bar\Phi}=0.
\end{equation}

The coupled equations so obtained are: 

\begin{eqnarray}\label{sigma1}
	&&\frac{\partial \Omega}{\partial \sigma}= k_{0}\chi^{2}\sigma-4k_{1}\left( \sigma^{2}+\zeta^{2}
	+\delta^{2}\right)\sigma-2k_{2}\left( \sigma^{3}+3\sigma\delta^{2}\right)
	-2k_{3}\chi\sigma\zeta \nonumber\\
	&-&\frac{d}{3} \chi^{4} \bigg (\frac{2\sigma}{\sigma^{2}-\delta^{2}}\bigg )
	+\left( \frac{\chi}{\chi_{0}}\right) ^{2}h_x- 
	\left(\frac{\chi}{\chi_0}\right)^2m_\omega\omega^2
	\frac{\partial m_\omega}{\partial\sigma}-\left(\frac{\chi}{\chi_0}\right)^2m_\rho\rho^2 
	\frac{\partial m_\rho}{\partial\sigma} \nonumber\\
	&-&\sum_{i=u,d} g_{\sigma}^i\left(\rho_{i}^{s} - \frac{N_cm_i^*}{8\pi^2}\left[-4\lambda\sqrt{\lambda^2+m_i^{*2}} + m_i^{*2}\left(1 + 4 {\rm ln}\frac{\lambda+\sqrt{\lambda^2+m_i^{*2}}}{\Lambda_0}\right)\right]\right) = 0 ,
\end{eqnarray}
\begin{eqnarray}
	&&\frac{\partial \Omega}{\partial \zeta}= k_{0}\chi^{2}\zeta-4k_{1}\left( \sigma^{2}+\zeta^{2}+\delta^{2}\right)
	\zeta-4k_{2}\zeta^{3}-k_{3}\chi\left( \sigma^{2}-\delta^{2}\right)-\frac{d}{3}\frac{\chi^{4}}{{\zeta}}\nonumber\\
	&+&\left(\frac{\chi}{\chi_{0}} \right)
	^{2}h_y-\left(\frac{\chi}{\chi_0}\right)^2m_\phi\phi^2 
	\frac{\partial m_\phi}{\partial\zeta} \nonumber \\
	&-& g_{\zeta}^s\left(\rho_{s}^{s} - \frac{N_cm_s^*}{8\pi^2}\left[-4\lambda\sqrt{\lambda^2+m_s^{*2}} + m_s^{*2}\left(1 + 4 {\rm ln}\frac{\lambda+\sqrt{\lambda^2+m_s^{*2}}}{\Lambda_0}\right)\right]\right) = 0 ,
	\label{zeta}
\end{eqnarray}
\begin{eqnarray}
	&&\frac{\partial \Omega}{\partial \delta}=k_{0}\chi^{2}\delta-4k_{1}\left( \sigma^{2}+\zeta^{2}+\delta^{2}\right)
	\delta-2k_{2}\left( \delta^{3}+3\sigma^{2}\delta\right) +\mathrm{2k_{3}\chi\delta
		\zeta} +  \frac{2}{3} d \chi^4 \left( \frac{\delta}{\sigma^{2}-\delta^{2}}\right) \nonumber \\
	&-&\sum_{i=u,d} g_{\delta}^i\left(\rho_{i}^{s} - \frac{N_cm_i^*}{8\pi^2}\left[-4\lambda\sqrt{\lambda^2+m_i^{*2}} + m_i^{*2}\left(1 + 4 {\rm ln}\frac{\lambda+\sqrt{\lambda^2+m_i^{*2}}}{\Lambda_0}\right)\right]\right) = 0 ,
	\label{delta}
\end{eqnarray}
\begin{eqnarray}
	&&\frac{\partial \Omega}{\partial \chi}=\mathrm{k_{0}\chi} \left( \sigma^{2}+\zeta^{2}+\delta^{2}\right)-k_{3}
	\left( \sigma^{2}-\delta^{2}\right)\zeta + \chi^{3}\left[1
	+{\rm {ln}}\left( \frac{\chi^{4}}{\chi_{0}^{4}}\right)  \right]
	+(4k_{4}-d)\chi^{3}	\nonumber\\
	&-&\frac{4}{3} d \chi^{3} {\rm {ln}} \Bigg ( \bigg (\frac{\left( \sigma^{2}
		-\delta^{2}\right) \zeta}{\sigma_{0}^{2}\zeta_{0}} \bigg )
	\bigg (\frac{\chi}{\mathrm{\chi_0}}\bigg)^3 \Bigg )+
	\frac{2\chi}{\chi_{0}^{2}}\left[ h_x\sigma +h_y \zeta\right] \nonumber\\
	&-& \frac{\chi}{{\chi^2}_0}({m_{\omega}}^2 \omega^2+{m_{\rho}}^2\rho^2+{m_{\phi}}^2\phi^2)  = 0,
	\label{chi}
\end{eqnarray}
\begin{eqnarray}
	\frac{\partial \Omega}{\partial \omega}=\frac{\chi^2}{\chi_0^2}m_\omega^2\omega+4g_4\omega^3+12g_4\omega\rho^2
	&-&\sum_{i=u,d}g_\omega^i\rho_{i}=0,
	\label{omega} 
\end{eqnarray}
\begin{eqnarray}
	\frac{\partial \Omega}{\partial \rho}=\frac{\chi^2}{\chi_0^2}m_\rho^2\rho+4g_4\rho^3+12g_4\omega^2\rho&-&
	\sum_{i=u,d}g_\rho^i\rho_{i}=0, 
	\label{rho} 
\end{eqnarray}
\begin{eqnarray}
	\frac{\partial \Omega}{\partial \phi}=\frac{\chi^2}{\chi_0^2}m_\phi^2\phi+8g_4\phi^3&-&
	g_\phi^s\rho_{s}=0,
	\label{phi}  
\end{eqnarray}
\begin{eqnarray}
	\hspace*{0.4cm} 
	\frac{\partial \Omega}{\partial \Phi} =\bigg[\frac{-a(T)\bar{\Phi}}{2}-\frac{6b(T)
		(\bar{\Phi}-2{\Phi}^2+{\bar{\Phi}}^2\Phi)
	}{1-6\bar{\Phi}\Phi+4(\bar{\Phi}^3+\Phi^3)-3(\bar{\Phi}\Phi)^2}\bigg]T^4
	-\sum_{i=u,d,s}\frac{2k_BTN_C}{(2\pi)^3}
	\nonumber\\
	\int_0^\infty d^3k 
	\bigg[\frac{e^{-(E_i^*(k)-{\mu_i}^{*})/k_BT}}{\left(1+e^{-3(E_i^*(k)-{\mu_i}^{*})/k_BT}+3\Phi e^{-(E_i^*(k)-{\mu_i}^{*})/k_BT}
		+3\bar{\Phi}e^{-2(E_i^*(k)-{\mu_i}^{*})/k_BT}\right)}
	\nonumber\\
	+\frac{e^{-2(E_i^*(k)+{\mu_i}^{*})/k_BT}}{\left(1+e^{-3(E_i^*(k)+{\mu_i}^{*})/k_BT}
		+3\bar{\Phi} e^{-(E_i^*(k)+{\mu_i}^{*})/k_BT}+3\Phi e^{-2(E_i^*(k)+{\mu_i}^{*})/k_BT}\right)}\bigg] \nonumber \\=0,
	\label{Polyakov} 
\end{eqnarray}
and
\begin{eqnarray}
	\frac{\partial \Omega}{\partial \bar{\Phi}} =\bigg[\frac{-a(T)\Phi}{2}-\frac{6b(T)
		(\Phi-2{\bar{\Phi}}^2+{\Phi}^2\bar{\Phi})
	}{\mathrm{1-6\bar{\Phi}\Phi+4(\bar{\Phi}^3+\Phi^3)-3(\bar{\Phi}\Phi)^2}}\bigg]T^4
	-\sum_{i=u,d,s}\frac{2k_BTN_C}{(2\pi)^3}
	\nonumber\\
	\int_0^\infty d^3k\ \bigg[\frac{e^{-2(E_i^*(k)-{\mu_i}^{*})/k_BT}}{\left(1+e^{-3(E_i^*(k)-{\mu_i}^{*})/k_BT}+3\Phi e^{-(E_i^*(k)-{\mu_i}^{*})/k_BT}
		+3\bar{\Phi}e^{-2(E_i^*(k)-{\mu_i}^{*})/k_BT}\right)}
	\nonumber\\
	+\frac{e^{-(E_i^*(k)+{\mu_i}^{*})/k_BT}}{\left(1+e^{-3(E_i^*(k)+{\mu_i}^{*})/k_BT}
		+3\bar{\Phi} e^{-(E_i^*(k)+{\mu_i}^{*})/k_BT}+3\Phi e^{-2(E_i^*(k)+{\mu_i}^{*})/k_BT}\right)}\bigg]\nonumber\\=0. 
	\label{Polyakov conjugate} 
\end{eqnarray}

The scalar density $\rho_{i}^{s}$ and the number (vector) density $\rho_{i}$ of the quarks are defined as
\begin{align}
	\rho_{i}^{s} &= \gamma_{i}N_c\int_{\lambda}^{\infty}\frac{d^{3}k}{(2\pi)^{3}} 
	\frac{m_{i}^{*}}{E^{\ast}_i(k)} \Big(f_i(k)+\bar{f}_i(k)
	\Big),
	\label{rhos0}\\
	\rho_{i} &= \gamma_{i}N_c\int_{\lambda}^{\infty}\frac{d^{3}k}{(2\pi)^{3}}  
	\Big(f_i(k)-\bar{f}_i(k)
	\Big),
	\label{rhov0}
\end{align} 
respectively, where $f_{i}(k)$ and $\bar{f}_{i}(k)$ are the Fermi distribution functions for quark and anti-quark at finite temperature, respectively, and are given by
\begin{eqnarray}
	f_{i}&=&\frac{\Phi e^{-E^-}+2\bar{\Phi} e^{-2E^-}+e^{-3E^-}}
	{[1+3\Phi e^{-E^-}+3\bar{\Phi} e^{-2E^-}+e^{-3E^-}]} ,
	\label{distribution}
\end{eqnarray}
\begin{eqnarray}
	\bar{f}_{i}&=&\frac{\bar{\Phi} e^{-E^+}+2\Phi e^{-2E^+}+e^{-3E^+}}
	{[1+3\bar{\Phi} e^{-E^+}+3\Phi e^{-2E^+}+e^{-3E^+}]}.
	\label{adistribution}
\end{eqnarray} 
Furthermore, the list of parameters of the PCQMF model used in the present work is summarized in Table \ref{tab:1}. The parameters are fitted to produce the correct vacuum masses of the $\pi,K,\sigma$ mesons and average masses of $\eta$ and $\eta^{'}$ \cite{wang2003}. The baryon, isospin, and strangeness chemical potentials for asymmetric quark matter are defined through the relations $\mu_B = \frac{3}{2}(\mu_u+\mu_d),\mu_I = \frac{1}{2}(\mu_u-\mu_d)$, and $\mu_S = \frac{1}{2}(\mu_u+\mu_d-2\mu_s)$, respectively. Here, $\mu_u,\mu_d,\mu_s$ are the chemical potentials of the up, down, and strange quarks, respectively.

\subsection{\label{transport}Transport coefficients}
In the present manuscript, we investigate transport coefficients $\eta,\zeta_b,\sigma_{el},$ and $\kappa$ within the context of the PCQMF model and examine their modifications due to the finite size consideration of the medium.
The transport coefficients for a system in a hydrodynamical regime can be calculated by employing the Kubo formalism \cite{kubo,ghosh2019} where the relaxation time is assumed to be smaller than the lifetime of the system. Within this assumption, the departure of the system from equilibrium contains only linear terms in the spatial and temporal gradients of thermodynamic parameters, e.g. $T,v$, etc. The expressions of the transport coefficients obtained using such a formalism are exactly the same as those derived within a quasiparticle approach in the kinetic theory under the relaxation time approximation (RTA) \cite{sasaki2010,deb,chakraborty,arnold}. In this work, we shall follow the kinetic theory approach to calculate the expressions for the transport coefficients. The transport coefficients $\eta, \zeta_b, \sigma_{el},\kappa$ can be determined using the Boltzmann transport equation, which can be written in the relaxation time approximation (RTA) as \cite{patra}
\begin{equation}
	\frac{\partial f_i^{'}}{\partial t} + \frac{\Vec{k}}{m_i^*}.\Vec{\nabla} f_i^{'} + \Vec{F}.\frac{\partial f_i^{'}}{\partial \Vec{k}} = \left(\frac{\partial f_i^{'}}{\partial t}\right)_{coll} = -\frac{1}{\tau}(f_{i}^{'}(\Vec{x},\Vec{k},t)-f_{i}(\Vec{x},\Vec{k},t)),
\end{equation}
where $\vec F$ is the force field acting on the particles in the medium, $f_i$ is the local equilibrium distribution of quarks given in Eq. (\ref{distribution}) and $f^{'}_{i}$ is the non-equilibrium distribution function for the quark. To study the transport coefficients, we are interested in small departures of the distribution function from the equilibrium in the hydrodynamic limit, and we define
\begin{equation}
	\delta f_{i}(\vec x,\vec k,t) = f_{i}^{'}(\vec x,\vec k,t) - f_{i}(\vec x,\vec k,t).
\end{equation}
\par The transport coefficients of the relativistic fluid are calculated with the help of necessary macroscopic quantities: the energy-momentum tensor $T_{\mu\nu}$ describes the energy and momentum density of the system, the four-dimensional quark baryon current $Q^{\mu}$ represents the flow of baryon charge, and the electric current $J^{\mu}$ captures the flow of electric charge. 
Considering the three flavour quark and antiquark, these macroscopic quantities are expressed in the kinetic theory as \cite{islam}
\begin{eqnarray}
	T_{\mu\nu} &=& 2N_cN_f\int\frac{d^3 k}{(2\pi)^3}\frac{k^{\mu}k^{\nu}}{E_i^*}(f^{'}_i + \bar{f^{'}_{i}}),
\end{eqnarray}
\begin{eqnarray}
	Q^{\mu}&=& 2N_cN_f\int \frac{d^3 k}{(2\pi)^3}\frac{k^{\mu}}{E_i^*}(f^{'}_i - \bar{f^{'}_{i}}),
\end{eqnarray}
\begin{eqnarray}
	J^{\mu}&=& 2N_c\sum_{i=u,d,s}\int\frac{d^3k}{(2\pi)^3}\frac{k^{\mu}}{E_i^*}(e_if^{'}_i + e_{\bar i}\bar{f^{'}_{i}}),
\end{eqnarray}  
where $N_c = N_f = 3$ are the colour and flavour degeneracy, respectively; $e_{u,\bar u} = \pm 2/3$, $e_{d/s,\bar d/\bar s} = \mp 1/3$; $k^{\mu} = (E_i^*, \vec k)$ is the particle four-momentum.

The shear viscosity $\eta$ and the bulk viscosity $\zeta_b$ can be calculated by splitting the ideal and dissipative parts of $T_{\mu\nu}$, $ Q^{\mu}$ is related to the thermal conductivity $\kappa$  while $\sigma_{el}$ can be estimated from the microscopic version of Ohm's law. The specific mathematical expressions are given as \cite{gavin,deb,islam}
\begin{align}
	T^{\mu\nu} &= T_0^{\mu\nu} + \Delta T^{\mu\nu}, \\ 
	Q^{\mu} &= Q_0^{\mu} + \Delta Q^{\mu}, \\
	\vec{J} &= \sigma_{el}\vec{E},
\end{align}
where the dissipative parts are given by \cite{deb}
\begin{align}
	\Delta T^{\mu\nu} &= \eta\left(D^{\mu}D^{\nu} + D^{\nu}D^{\mu} + \frac{2}{3}\Delta^{\mu\nu}\delta_{\alpha}u^{\alpha}\right) - \zeta_b\delta_{\alpha}u^{\alpha}, \\
	\Delta Q^{\mu} &= \kappa \frac{T^2}{h}\Delta^{\mu\nu}D_{\nu}\left(\frac{\mu}{T}\right).
\end{align}
A detailed derivation of the transport coefficients, $\eta, \zeta_b, \sigma_{el}$ and $\kappa$ obtained through the kinetic theory in the RTA can be seen in Refs. \cite{plumari,hosoya,chakraborty,kadam2018} and through the one loop diagram approximation in the quasiparticle Kubo approach in Refs. \cite{ghosh2014,ghosh2019,fernandez}. These expressions are outlined below \cite{saha2018,ghosh2019}

\begin{eqnarray}
	\eta&=&\frac{2N_{c}}{15T}\sum_{i=u,d,s}\int\frac{{\rm d}^3k}{(2\pi)^3}\tau\left(\frac{k^{2}}{E_i^{*}}\right)^{2}[f_{i}(1-f_{i})+\bar{f}_{i}(1-\bar{f}_{i})],
	\label{eta}
\end{eqnarray}
\begin{eqnarray}
	\zeta_b&=&\frac{2N_{c}}{T}\sum_{i=u,d,s}\int\frac{{\rm d}^3k}{(2\pi)^3}\tau\frac{1}{E_i^{*2}}\left[\left(\frac{1}{3}-c_{s}^{2}\right)k^{2}-c_{s}^{2}m_i^{*2}+c_{s}^{2}m_i^{*}T\frac{dm_i^{*}}{dT}\right]^{2} \nonumber \\ 
	&&\left[f_{i}(1-f_{i})+\bar{f}_{i}(1-\bar{f}_{i})\right], 
	\label{zeta}
\end{eqnarray}
\begin{eqnarray}
	\sigma_{el}&=&\frac{2N_{c}}{3T}\sum_{i=u,d,s}e_i^{2}\int\frac{{\rm d}^3k}{(2\pi)^3}\tau\left(\frac{k}{E_i^{*}}\right)^{2}[f_{i}(1-f_{i})+\bar{f}_{i}(1-\bar{f}_{i})],
	\label{sigma}
\end{eqnarray}
\begin{eqnarray}
	\kappa &=& \frac{2N_c}{3T^2}\sum_{i=u,d,s}\int\frac{{\rm d}^3k}{(2\pi)^3}\tau\left(\frac{k}{E_i^*}\right)^2[(E_i^*-h)^2f_i(1-f_i)+(E_i^*+h)^2 \nonumber \\
	&&\bar{f}_i(1-\bar{f}_i)].
	\label{kappa}
\end{eqnarray}

Notably, $f_i$ and $\bar{f}_i$ represent equilibrium distribution functions for quarks and antiquarks given in Eqs. (\ref{distribution}) and (\ref{adistribution}), respectively. 
 The term $c_s^2$ is the squared speed of sound, which at constant entropy is defined by $c_s^2 = \left(\frac{\partial p}{\partial \epsilon}\right)_{s} = \frac{s}{c_{v}}$, where the pressure $p = -\Omega$, the energy $\epsilon = \Omega+\sum_{i} {\mu_i}^{*} \rho_i+Ts$, the entropy $s= -\frac{\partial\Omega}{\partial T}$, and the specific heat $c_{v} = \left(\frac{\partial \epsilon}{\partial T}\right)_V.$ The heat function $h = (\epsilon+p)/\rho$, where $\rho$ is the net quark density. This quantity diverges at $\mu=0$ where $\rho$ vanishes.  

In order to calculate the transport coefficients, given by Eqs. (\ref{eta}) - (\ref{kappa}), we need to know the relaxation time $\tau$, which refers to the timescale over which the collisions cause the distribution function to relax to an equilibrium state and is given as \cite{hosoya}
\begin{equation}
	\tau = \frac{1}{5.1T\alpha_S^2\log(\frac{1}{\alpha_S})(1+0.12(2N_f+1))}.
	\label{tau}
\end{equation}
In above, $\alpha_S$ is the strong coupling constant dependent on temperature $T$ and quark chemical potential $\mu$ and is written as \cite{bannur, zhu}
\begin{equation}
	\alpha_S(T,\mu)=\frac{6\pi}{(33-2N_f)\log\left(\frac{T}{\Lambda_T}\sqrt{1+(\frac{\mu}{\pi T})^2}\right)}\left(1-\frac{3(153-19N_f)}{(33-2N_f)^2}\frac{\log\left(2\log\frac{T}{\Lambda_T}\sqrt{1+(\frac{\mu}{\pi T})^2}\right)}{\log\left(\frac{T}{\Lambda_T}\sqrt{1+(\frac{\mu}{\pi T})^2}\right)}\right),
	\label{alpha}
\end{equation}
with $\Lambda_T=70$ MeV \cite{zhu}. Notably, Eq. (\ref{tau}) is valid for the massless quarks. However, as discussed in Ref. \cite{berre}, the effect of the mass of the quarks on the relaxation time is small. In the present work, we have used the relaxation time given in Eq. (\ref{tau}) to calculate the transport coefficients.

\begin{table}[h]
	\scriptsize{
		\centering
		\begin{tabular}{|c|c|c|c|c|c|c|c|c|c|}
			\hline
			$k_0$           & $k_1$          & $k_2$          & $k_3$         & $k_4$         & $g_s$         & $\rm{g_v}$          & $\rm{g_4}$           & $d$          & $\rho_0$(fm$^{-3}$)                            \\ \hline
			0.2002                 & 2.3882                & -19.4995              & -4.7334              & -0.06              & 4.76               & 4               & 37.5                 & 0.002               & 0.15                                  \\ \hline
			$\sigma_0$ (MeV) & $\zeta_0$(MeV)  & $\chi_0$(MeV)   & $m_\pi$(MeV)  & $f_\pi$(MeV)  & $m_K$(MeV)    & $f_K$(MeV)     & $m_\omega$(MeV) & $m_\phi$(MeV)  & $m_\rho$( MeV)                   \\ \hline
			-93                  & -95.47              & 254.6               & 139                & 93                 & 496                & 115                 & 783                  & 1020                & 783                                   \\ \hline
			$g_{\sigma}^u$  & $g_{\sigma}^d$ & $g_{\sigma}^s$ & $g_{\zeta}^u$ & $g_{\zeta}^d$ & $g_{\zeta}^s$ & $g_{\delta}^u$ & $g_{\delta}^d$  & $g_{\delta}^s$ & $g^u_{\omega}$ \\ \hline
			3.36                 & 3.36                & 0                   & 0                  & 0                  & 4.76               & 3.36                & -3.36                & 0                   &      3.86                             \\ \hline
			$g^d_{\omega}$ & $g^s_{\omega}$ & $g^u_{\phi}$ &  $g^d_{\phi}$ & $g^s_{\phi}$  & $g^u_{\rho}$   & $g^d_{\rho}$  &   $g^s_{\rho}$  & $\Lambda_0$ (MeV)  &                             \\ \hline
			3.86       &        0         &           0         &    0               &           5.46        &      3.86          &    -3.86             &          0       &               600      &                                \\ \hline
		\end{tabular}
		\caption{The list of parameters used in the present work.}
		\label{tab:1}
	}
\end{table}



\section{\label{results}Results and Discussions}
In this section, we will discuss the effects of finite volume on the transport coefficients of quark matter within the framework of the Polyakov chiral SU(3) quark mean-field model. As discussed earlier, the effects of finite volume, characterized by the length of the cubic volume $R$, come into the picture by substituting $k=0$ with $k=\lambda=\pi/R$ in the lower limit of the integration in Eq. (\ref{tpdpcqmf}). This causes a modification in the values of scalar fields $\sigma,\zeta,$ and $\delta$, which in turn modifies the effective mass, $m_i^*$, of quarks through Eq. (\ref{mbeff}). Additionally, the lower momentum cutoff is also applied to the equations of the transport coefficients, $\eta, \zeta_b, \sigma_{el}$ and $\kappa$ given in Eqs. (\ref{eta}) - (\ref{kappa}). 
We have also implemented the finite size to the fermion vacuum term presented in Eq. (\ref{omegavac}) and examined its impact on the effective quark masses, thermodynamic properties, and transport coefficients. Along with the finite size effects, we have considered the influence of polynomial and logarithmic forms of the Polyakov loop potentials with quark back reaction,
defined using Eqs. (\ref{polypoly}) and (\ref{log}), on the transport properties
. 

\par
\begin{figure*}
\centering
\begin{minipage}[c]{0.98\textwidth}
(a)\includegraphics[scale=0.5]{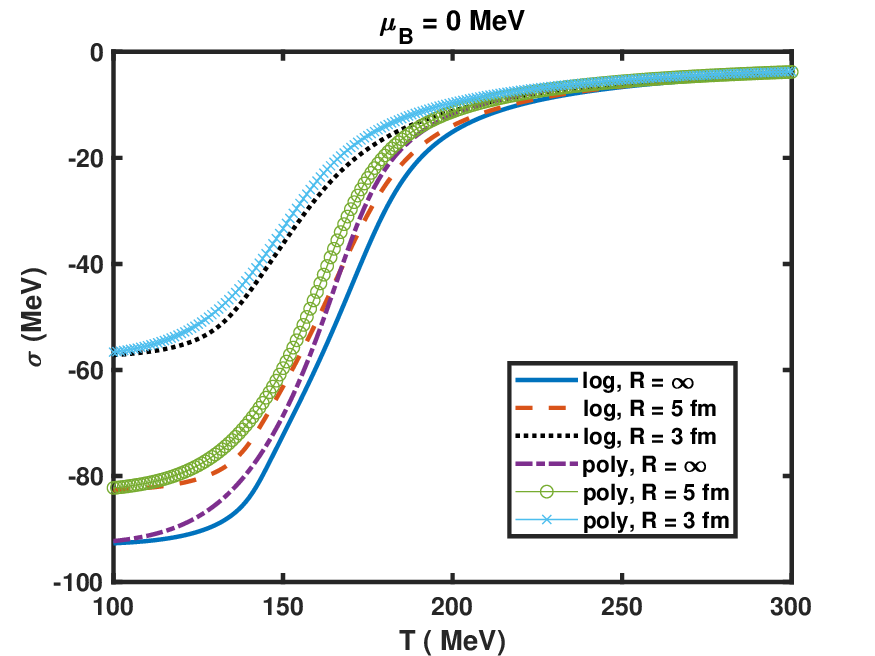}
\hspace{0.03cm}
(b)\includegraphics[scale = 0.5]{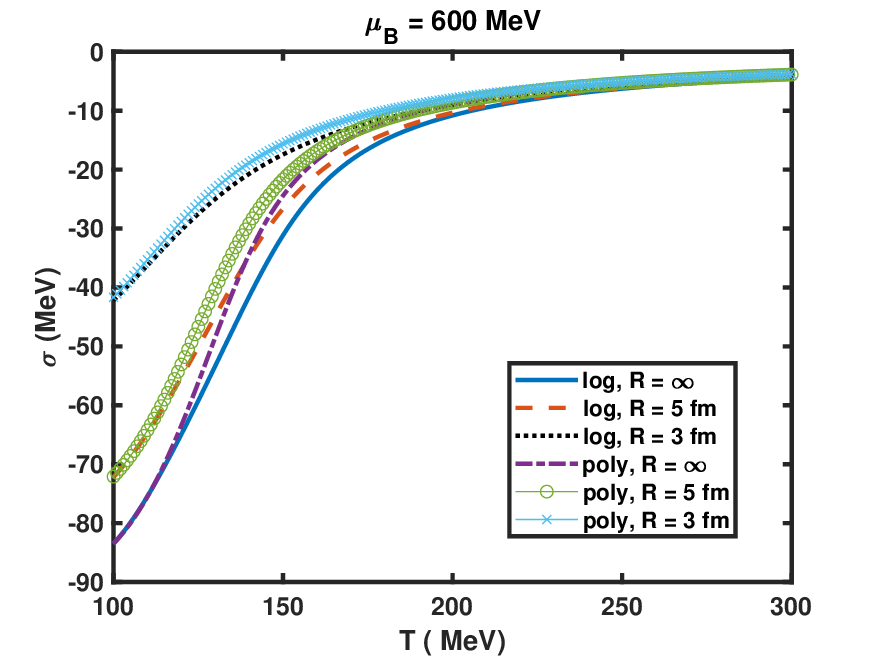}
\hspace{0.03cm}
(c)\includegraphics[width=7.4cm]{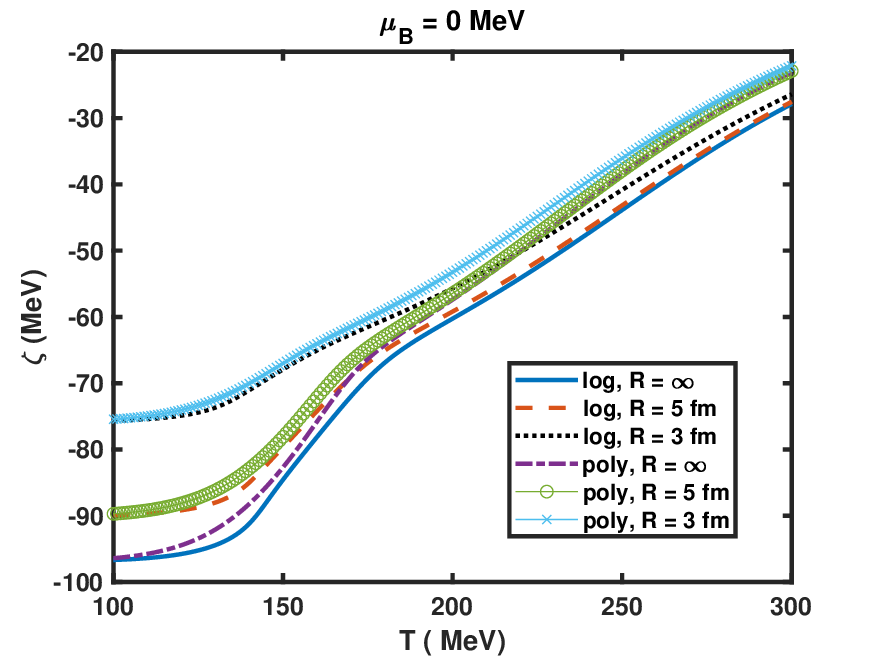}
\hspace{0.03cm}
(d)\includegraphics[width=7.4cm]{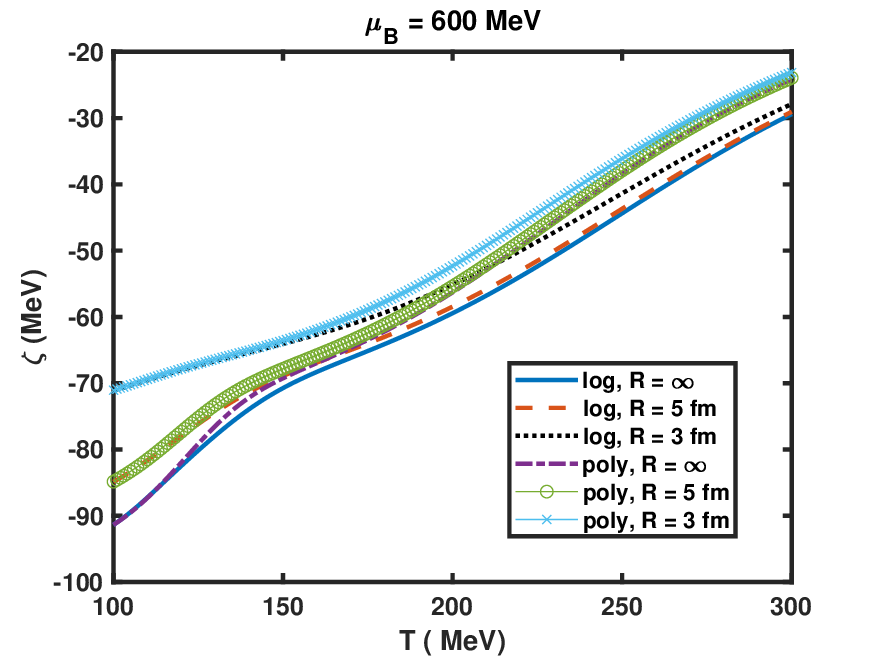}
\hspace{0.03cm}	
\end{minipage}
\caption{\label{fields} (Colour online) The scalar fields $\sigma$ and $\zeta$ are plotted as a function of temperature $T$ for lengths of cubic volume $R=\infty$, 5 fm, and $3$ fm, at baryon chemical potential $\mu_B = 0$ MeV [in subplots (a) and (c)], and baryon chemical potential $\mu_B = 600$ MeV, isospin chemical potential $\mu_I = -30$ MeV, and strangeness chemical potential $\mu_S = 125$ MeV [in subplots (b) and (d)], for both logarithmic Polyakov loop potential $\cal{U_L}$ and polynomial Polyakov loop potential $\cal{U_P}$.}
\end{figure*}

 We begin by discussing the influence of finite size effects on the scalar fields $\sigma$ and $\zeta$ 
which contribute to the medium modification of the effective quark masses. 
In Fig. \ref{fields}, we plot the scalar fields, $\sigma$ and $\zeta$ as a function of temperature $T$, at baryon chemical potentials $\mu_B = 0$  and $  600$ MeV, with $T_0^{glue}$ fixed at 200 MeV. The results are shown for the system sizes $R = \infty, 5$ fm, and $3$ fm, for the polynomial form of the Polyakov loop potential $\cal{U_P}$ as well as the logarithmic form $\cal{U_L}$.
We observe that the magnitude of the scalar fields decreases with an increase in the temperature of the medium. 
This reduction in the strength of the scalar fields at higher temperatures may signify the restoration of chiral symmetry. 
For a given value of temperature $T$, a decrease in the system size from $R=\infty$ to 3 fm results in a decrease in the magnitude of the scalar fields, $\sigma$ and $\zeta$. The influence of the finite size on the magnitude of the scalar fields appears to be more prominent in the low $T$ region as compared to higher $T$.
For example, at zero baryon chemical potential and temperature $T = 100$ MeV, for the Polyakov loop potential $\cal{U_L}$, the magnitude of the $\sigma$ field decreases from 92.7 MeV at $R = \infty$ to 57.2 MeV at $R = 3$ fm. This suggests the earlier restoration of the chiral symmetry in a system with finite volume. The subplots $(b)$ and $(d)$ of Fig. \ref{fields} show the variations of the scalar fields, $\sigma$ and $\zeta$, with temperature $T$, at $\mu_B = 600$ MeV and system sizes $R = \infty, 5$ fm, and $3$ fm. 
The isospin asymmetry in the medium is introduced through an isospin chemical potential $\mu_I = -30$ MeV, while the strangeness chemical potential is fixed at $ \mu_S = 125$ MeV.
The values of $\mu_I$ and $\mu_S$ chosen in the present study are motivated by Refs. \cite{liu,fu,tawfik17}, which explore the typical values of these chemical potentials in heavy-ion collision experiments.
We observe that at low temperatures, as the baryon chemical potential is increased from zero to a finite value, the magnitude of the scalar fields decreases drastically. 
For both Polyakov loop potential $\cal{U_L}$ and $\cal{U_P}$, at temperature $T = 150$ MeV and system size $R = \infty$, the magnitude of $\sigma$ field drops by about $57 \%$ as the value of $\mu_B$ is increased from 0 to 600 MeV.
This may signify the restoration of chiral symmetry at lower temperatures for increasing values of baryon chemical potential, which may be significant for future experimental facilities such as FAIR and NICA aimed to explore the QCD phase diagram at higher baryon chemical potential. At $\mu_B = 600$ MeV and $T= 150$ MeV, the magnitude of the $\sigma$ field is decreased by about $44\%$ when the system size changes from $R=\infty$ to $3$ fm. 

\begin{figure*}
\centering
\begin{minipage}[c]{0.98\textwidth}
(a)\includegraphics[width=7.4cm]{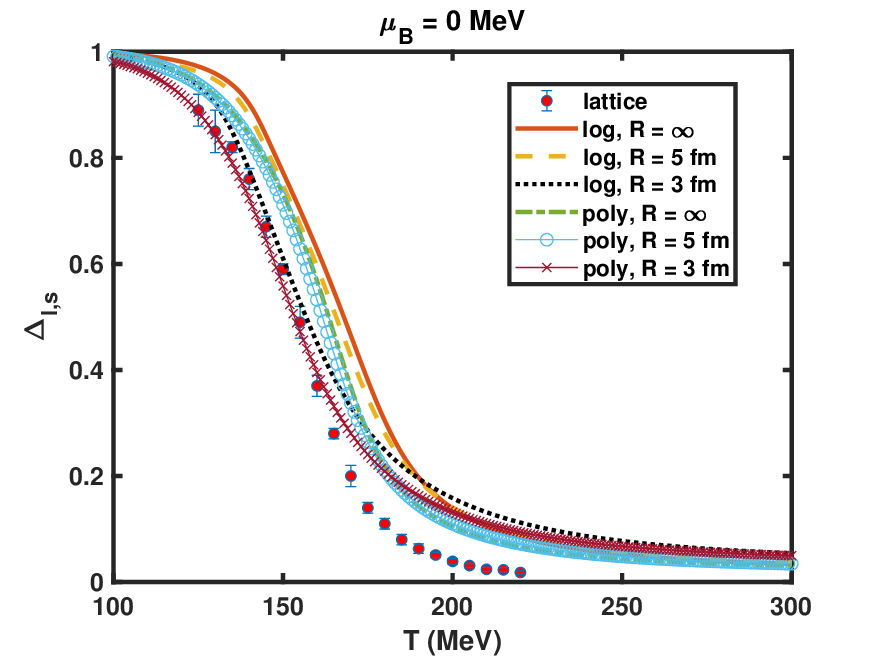}
\hspace{0.03cm}
(b)\includegraphics[width=7.4cm]{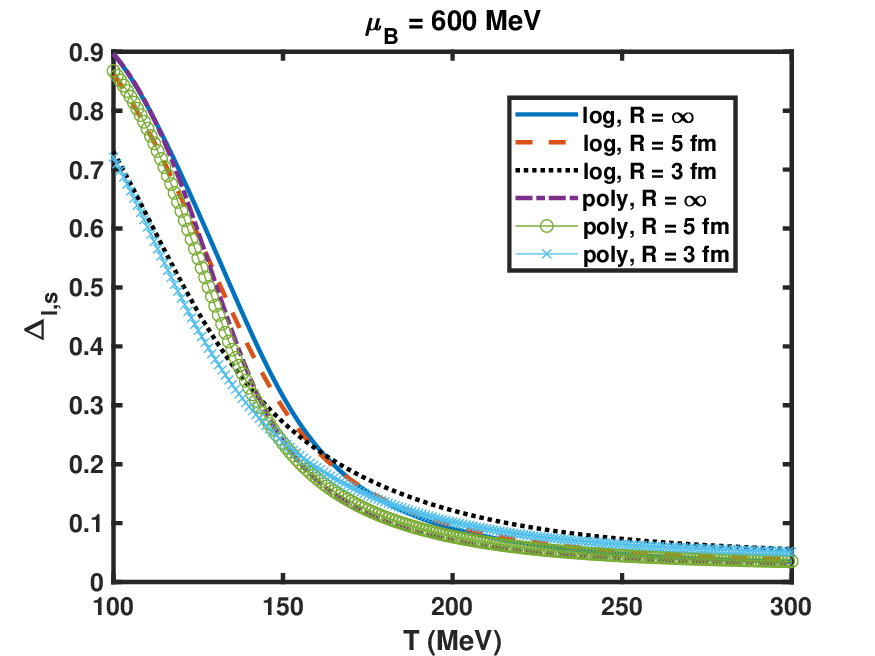}
\hspace{0.03cm}	
(c)\includegraphics[width=7.4cm]{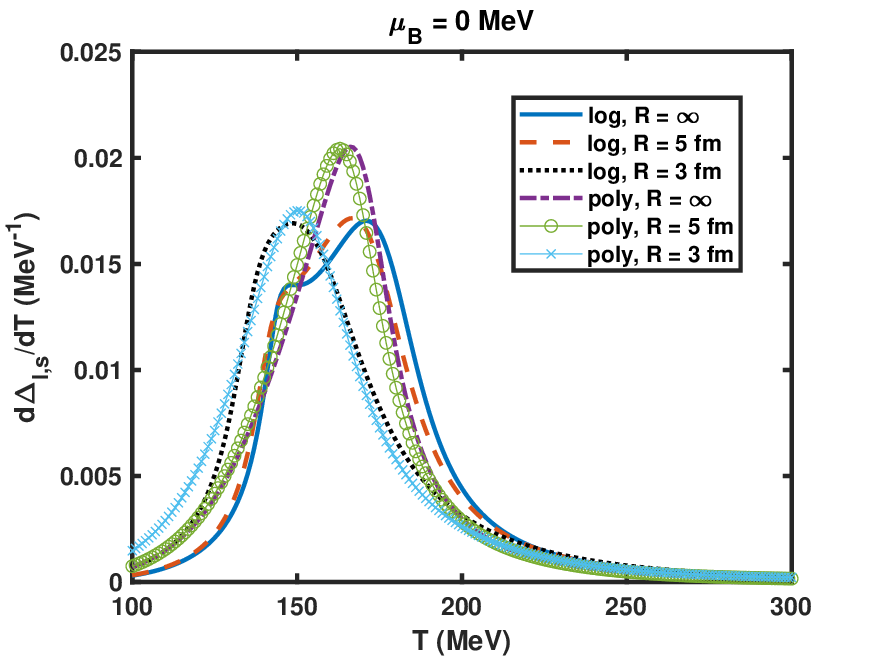}
\hspace{0.03cm}
(d)\includegraphics[width=7.4cm]{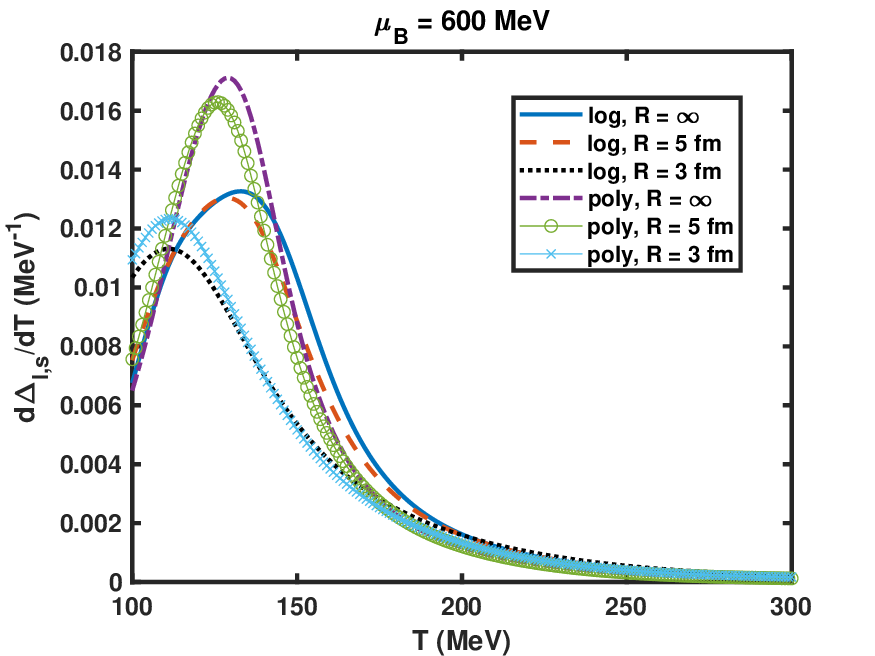}
\hspace{0.03cm}	
\end{minipage}
\caption{\label{fieldsd} (Colour online) The subtracted chiral condensate $\Delta_{l,s}$ is plotted as a function of temperature $T$ [in subplots (a) and (b)], and its derivative $d\Delta_{l,s}/dT$ [in subplots (b) and (d)], with $\mu_B = 0$ MeV and $\mu_B = 600$ MeV at system size $R = \infty$, 5 fm, and 3 fm, for Polyakov loop potentials $\cal{U_L}$ and $\cal{U_P}$.}
\end{figure*}

In the chiral limit, the quark condensate is expected to vanish completely at critical temperature $T_{\chi}$. At nonvanishing quark mass, the quark condensate requires some renormalization. Therefore, in order to compute the transition temperature of the chiral phase transition, we have calculated 
the subtracted chiral condensate defined as \cite{baza2009}
\begin{equation}
    \Delta_{l,s}(T) = \frac{\langle \bar{\psi}\psi\rangle_{l,T}-\frac{\hat{m}_l}{\hat{m}_s}\langle\bar{\psi}\psi\rangle_{s,T}}{\langle \bar{\psi}\psi\rangle_{l,0}-\frac{\hat{m}_l}{\hat{m}_s}\langle\bar{\psi}\psi\rangle_{s,0}}.
\end{equation}
In the PCQMF model, this term can be written as
\begin{equation}
    \Delta_{l,s}(T) = \frac{\sigma-\frac{h_x}{h_y}\zeta}{\sigma_0-\frac{h_x}{h_y}\zeta_0},
\end{equation}
where we have replaced the light quark condensate $\langle \bar{\psi}\psi\rangle_{l,T}$ and strange quark condensate $\langle \bar{\psi}\psi\rangle_{s,T}$ with the non-strange $\sigma$ and strange $\zeta$ field, respectively. As they are proportional to the chiral condensate $\langle\sigma_x\rangle$ and $\langle\sigma_y\rangle$, see, e.g., Ref. \cite{manisha}. In addition, the ratio of symmetry-breaking parameters $h_x$ and $h_y$ as they are directly related to the bare quark masses $\hat{m_l}$ and $\hat{m_s}$ \cite{schaefer2010}. In Fig. \ref{fieldsd}, we plot the temperature variations of $\Delta_{l,s}$ (subplots (a) and (b))and its temperature derivative $d\Delta_{l,s}/dT$ (subplots (c) and (d)), for the Polyakov loop potentials $\cal{U_L}$ and $\cal{U_P}$, at $R = \infty, 5$ fm, and $3$ fm and $\mu_B = 0$ and $600$ MeV. The result at vanishing baryon chemical potential $\mu_B$ is compared with the lattice data \cite{borsanyi2010}. 
The location of the chiral transition temperature $T_{\chi}$ can be obtained from the peak of the derivative of subtracted condensate $d\Delta_{l,s}/dT$. We observe that, with decreasing system size ($R=\infty\rightarrow$ 3 fm), the transition becomes smoother and is shifted towards lower temperature values, as can be seen in Fig. \ref{fieldsd} (c).
 This is consistent with the lattice calculations where the transition temperature is observed to shift to lower temperatures for smaller system size \cite{borsanyi2025,fodor}.
 At $\mu_B = 0$ MeV, the value of $T_{\chi}$ is lower ($T_{\chi}=166$ MeV) for $\cal U_P$ compared to $\cal U_L$ ($T_{\chi}=171$ MeV) at $R=\infty$. This implies that the chiral transition is sensitive to the particular form of the Polyakov loop potential considered. However, as the system size decreases from $R = \infty$ to 5 fm and $3$ fm, the value of $T_{\chi}$ shifts to 163 MeV (167 MeV) and 149 MeV (150 MeV) for $\cal{U_P}$ ($\cal{U_L}$), respectively. This shows that the chiral transition becomes less sensitive to the form of the Polyakov loop for finite size systems. From Fig. \ref{fieldsd}$(d)$, we observe that at baryon chemical potential $\mu_B=600$ MeV and system size $R=\infty$, the values of the transition temperature $T_{\chi}$ for the chiral phase transition is found to be 129 and 133 MeV, for the Polyakov loop potentials $\cal{U_P}$ and $\cal{U_L}$, respectively. This implies a shift in the chiral phase transition to lower temperatures with increasing baryon chemical potential. Decreasing the system size to $R = 3$ fm results in further shifting $T_{\chi}$ to 112 MeV and 111 MeV, for $\cal{U_L}$ and $\cal{U_P}$, respectively. 
In conclusion, we observe that the chiral phase transition is shifted towards lower temperatures when considering a system with finite size, and this shift is more for finite baryon chemical potential.

\par 

    \begin{figure*}
    \centering
    \begin{minipage}[c]{0.98\textwidth}
    (a)\includegraphics[width=7.1cm]{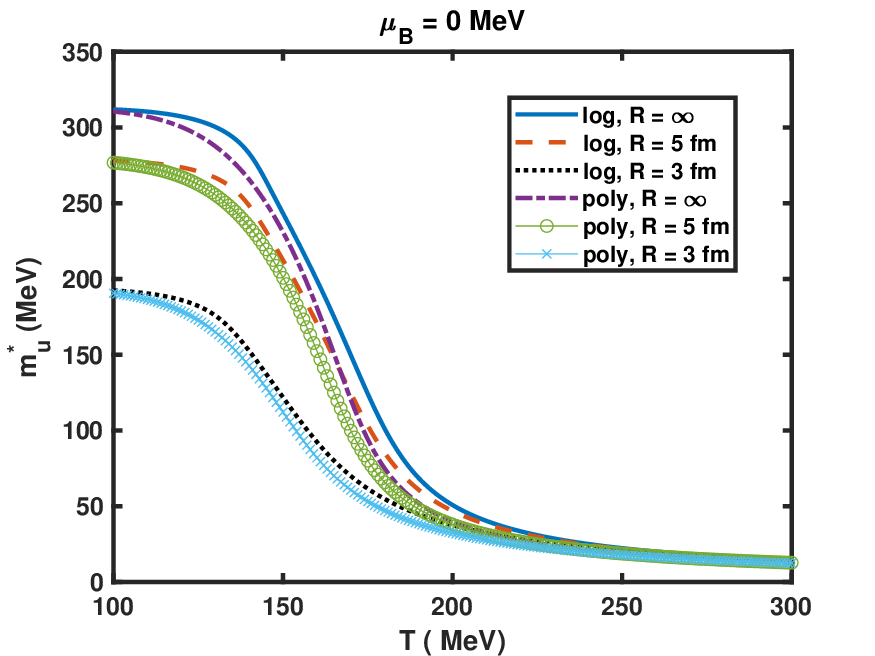}
    \hspace{0.03cm}
    (b)\includegraphics[width=7.1cm]{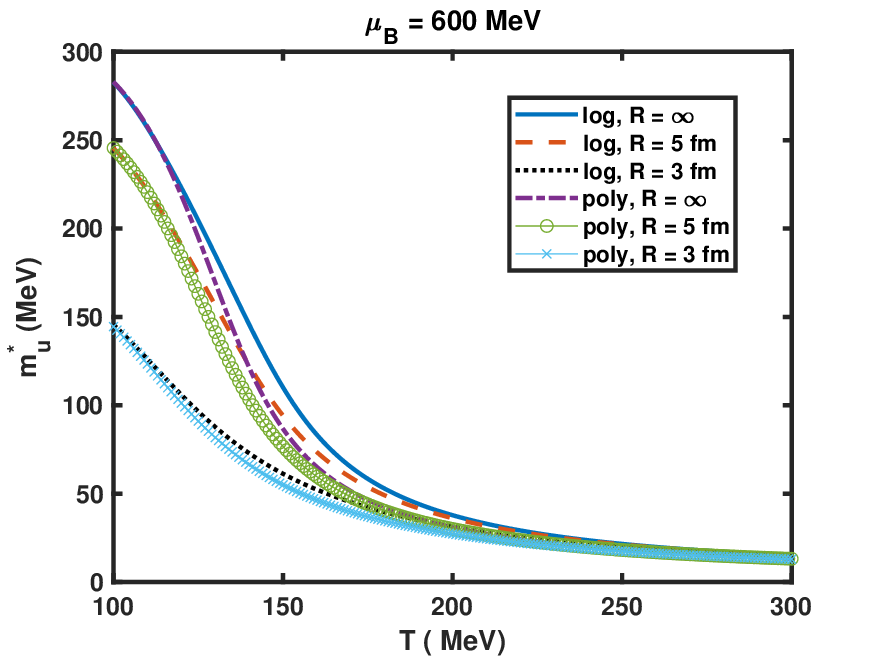}
    \hspace{0.03cm}	
    (c)\includegraphics[width=7.1cm]{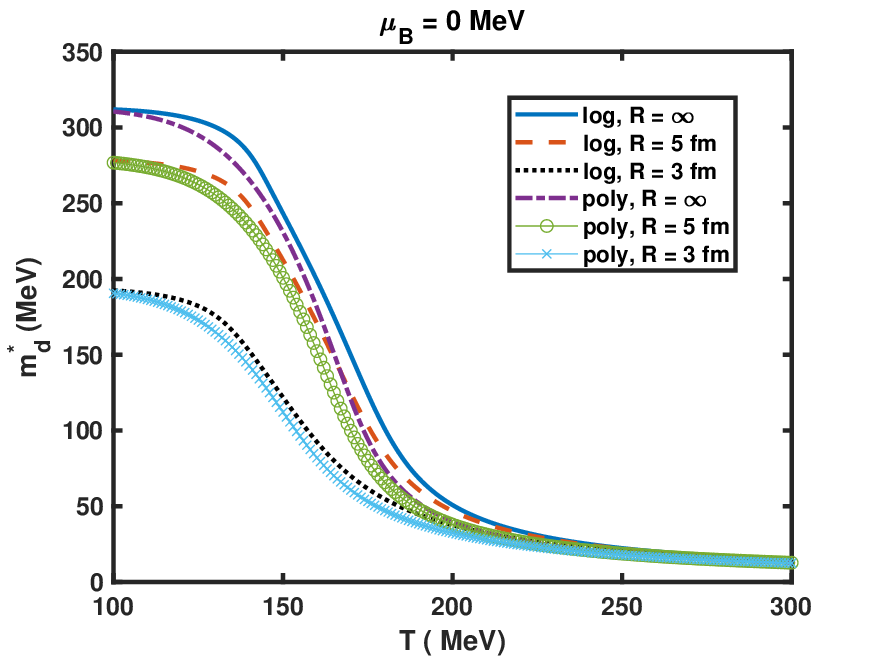}
    \hspace{0.03cm}
    (d)\includegraphics[width=7.1cm]{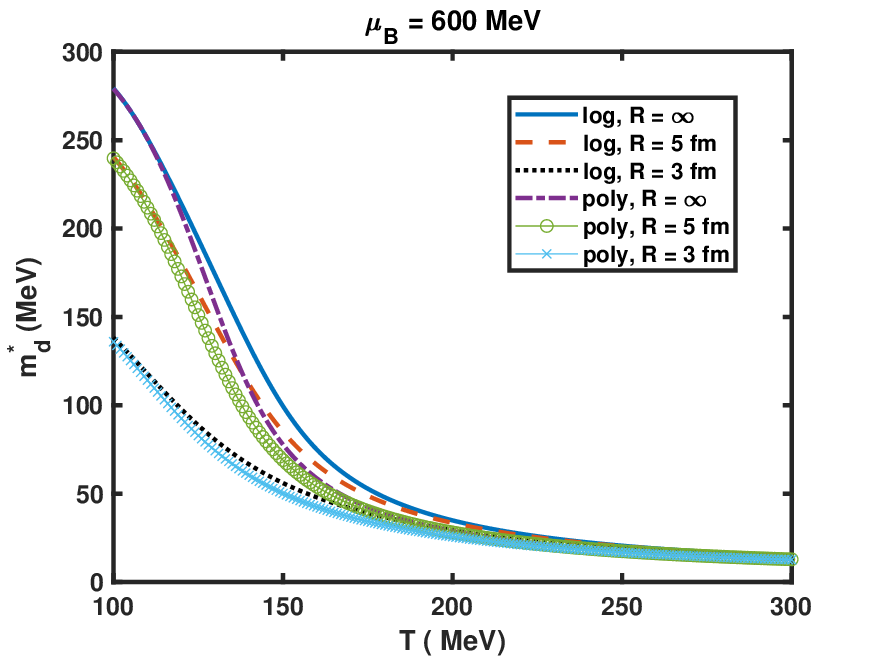}
    \hspace{0.03cm}	
    (e)\includegraphics[width=7.1cm]{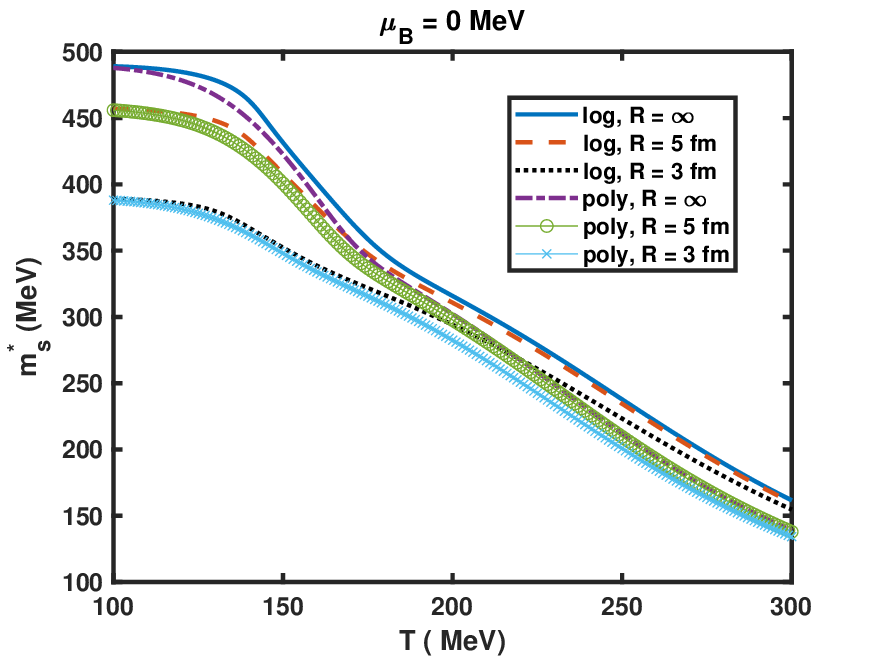}
    \hspace{0.03cm}
    (f)\includegraphics[width=7.1cm]{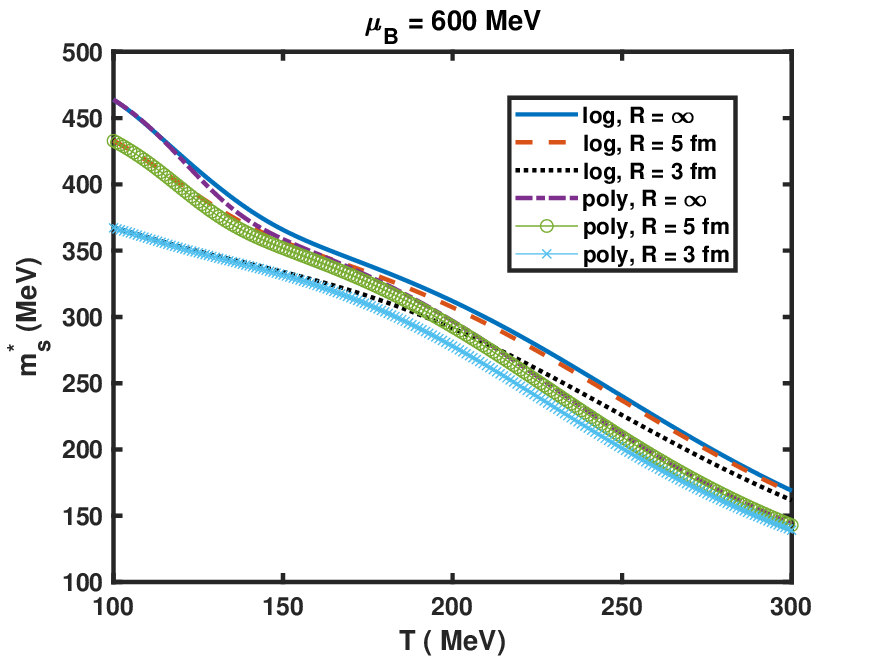}
    \hspace{0.03cm}
    \end{minipage}
    \caption{\label{mass} \footnotesize{(Colour online) The effective quark masses $m_u^*,m_d^*,$ and $m_s^*$ plotted as a function of temperature $T$ for lengths of cubic volume $R=\infty$, 5 fm, and $3$ fm, at baryon chemical potential $\mu_B = 0$ MeV [in subplots (a), (c), and (e)], and baryon chemical potential $\mu_B = 600$ MeV, isospin chemical potential $\mu_I = -30$ MeV, and strangeness chemical potential $\mu_S = 125$ MeV [in subplots (b),(d), and (f)], for both logarithmic Polyakov loop potential $\cal{U_L}$ and polynomial Polyakov loop potential $\cal{U_P}$.}}
    \end{figure*} 

\par In Fig. \ref{mass}, we explore the system size dependence in the effective quark masses $m^*_u, m^*_d,$ and $m^*_s$ as a function of temperature $T$, under the change of $R$ from $\infty$ to 5 fm and $3$ fm. We show the results at baryon chemical potentials $\mu_B = 0$ and $600$ MeV, for the Polyakov loop potentials $\cal{U_L}$ and $\cal{U_P}$. The temperature dependence of the effective masses of quarks is determined by the temperature dependent scalar fields $\sigma$, $\zeta$ and $\delta$, as is evident in Eq. (\ref{mbeff}). The effective quark masses are observed to decrease with increasing temperature $T$. This happens as the system may transition from a confined state of hadrons to a phase of deconfined QGP at higher temperatures. Interestingly, at zero baryon chemical potential (subplots (a), (c), and (e) of Fig. \ref{mass}), we observe a dramatic decrease in the effective quark masses as the system size is decreased from $R=\infty$ to $3$ fm. This decrease in the effective quark masses due to the decreasing size of the system is more prominent in the low-temperature regime and is negligible at high-temperature values. 
At $\mu_B=0$ MeV and $T=100$ MeV, when going from $R=\infty$ to 3 fm, we see that the effective masses of $u$ and $d$ quarks $m^*_{u,d}$ decrease by $38\%$, for both $\cal{U_L}$ and $\cal{U_P}$. For the effective mass of the strange quark $m^*_s$, we find this decrease to be about $20\%$.
This decrease in the effective masses of quarks for smaller size systems (low $R$) at low temperatures may result from the early chiral symmetry restoration, as discussed earlier in Figs. \ref{fields} and \ref{fieldsd}. As a result, the effective mass of quarks is lesser for smaller systems in that temperature regime.
A similar impact of finite system size is observed in the PNJL and PQM model with finite size consideration in the vacuum term \cite{saha2018,kovacs}. This result contrasts behaviour to the results previously obtained in the PCQMF model \cite{nisha}, where the quark masses increase for smaller systems in the high $T$ region. This may be due to the inclusion of the fermionic vacuum term in the thermodynamic potential in the present work. 

From subplots (b),(d), and (f) of Fig. \ref{mass}, we observe that increasing the baryon chemical potential to finite value $\mu_B= 600$ MeV leads to a reduction in the effective masses of quarks even at low $T$ values.
For example, for logarithmic Polyakov loop $\cal{U_L}$ at $R=\infty$ and $T=150$ MeV, increasing the baryon chemical potential $\mu_B$ from 0 to 600 MeV leads to a decrease in the effective masses of $u$ and $d$ quarks by about 54$\%$ and $59\%$, respectively. The difference in the percentage decrease in the effective masses of $u$ and $d$ quarks at finite $\mu_B$ is due to the introduction of finite isospin chemical potential $\mu_I = -30$ MeV. We find the effective mass of the strange quark to decrease only by $15\%$ when going from zero to finite baryon chemical potential. This may suggest that the chiral symmetry restoration of the strange quark happens at much higher $T$ and $\mu_B$ due to its much heavier mass as compared to the lighter $u$ and $d$ quarks. The reduction in the effective quark masses at finite $\mu_B$ is expected, as it depends on the scalar fields through Eq. (\ref{mbeff}), whose magnitude decreases when going from zero to finite baryon chemical potential, as discussed earlier. As for the impact of the finiteness of the system with finite $\mu_B$, we observe that at $T = 150$ MeV, decreasing the system size from $R = \infty$ to 3 fm leads to a decrease of $44\%$ (36$\%$) and $8\%$ ($7\%$) in $m^*_{u,d}$ and $m_s^*$ for $\cal{U_L}$ ($\cal{U_P}$). Again, the decrease in the effective quark masses for smaller systems may be due to the shifting of $T_{\chi}$ at lower temperatures, and the impact of finite size seems to be much less for the heavier $s$ quark compared to lighter $u$ and $d$ quarks. The impact of the finiteness of the system becomes less prominent at higher $T$ due to the restoration of chiral symmetry in that regime independent of the system size. Hence, it becomes acceptable to disregard the impact of finite size on the QGP generated in heavy ion experiments as long as it remains at a high temperature, well beyond the transition temperature, $T_{\chi}$. However, within the vicinity of $T_{\chi}$ and the non-perturbative (hadronic) temperature range, finite size effects remain significant, influencing not only quark masses but also various other quantities discussed later. This underscores the significance of finite size in the vicinity of the transition point. Similar observations were observed in Ref. \cite{saha2018}, where the authors used the PNJL model to study finite volume effects on the thermodynamic quantities of quark matter.

\begin{figure*}
    \centering
    \begin{minipage}[c]{0.98\textwidth}
    (a)\includegraphics[width=7.4cm]{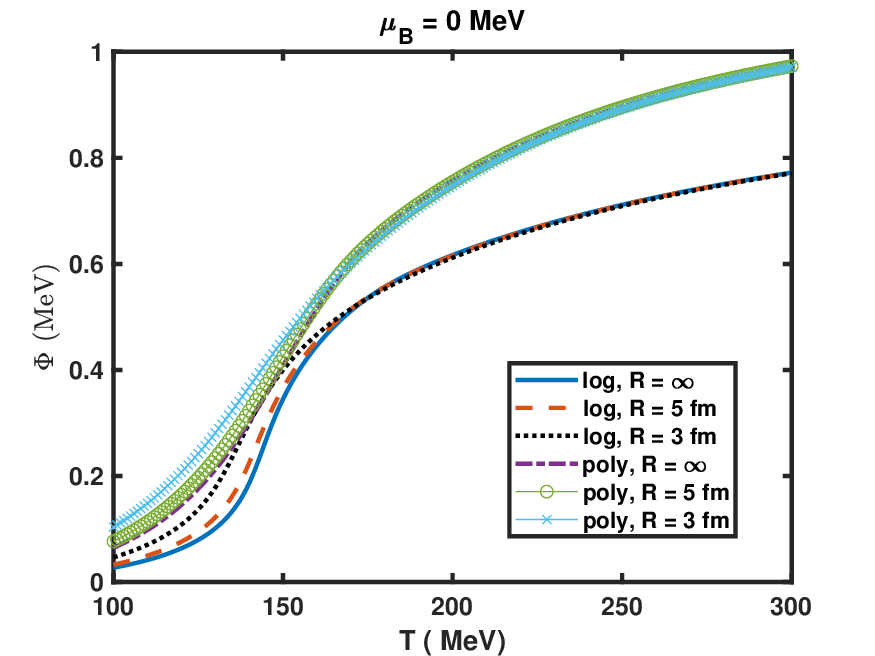}
    \hspace{0.03cm}
    (b)\includegraphics[width=7.4cm]{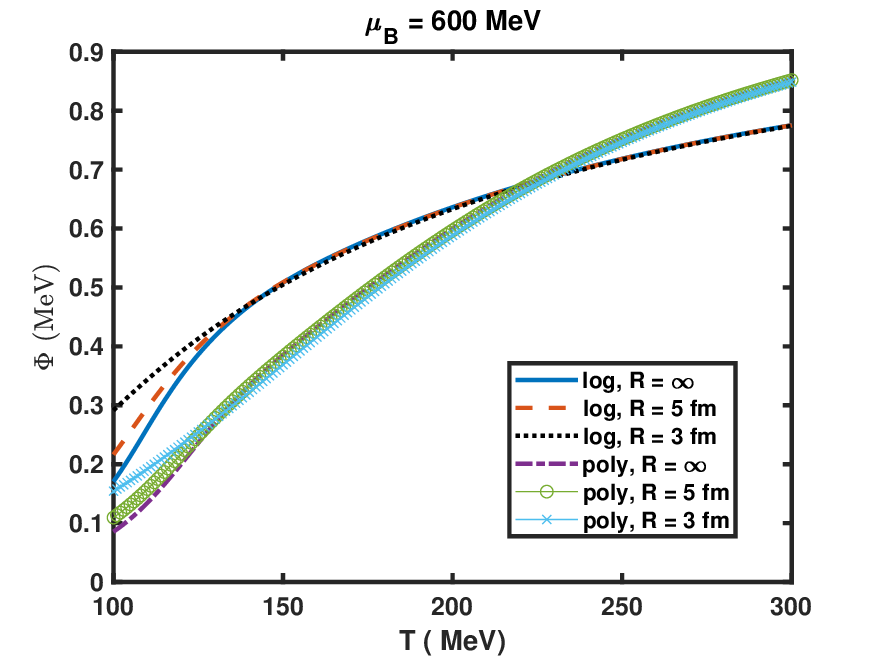}
    \hspace{0.03cm}	
    (c)\includegraphics[width=7.4cm]{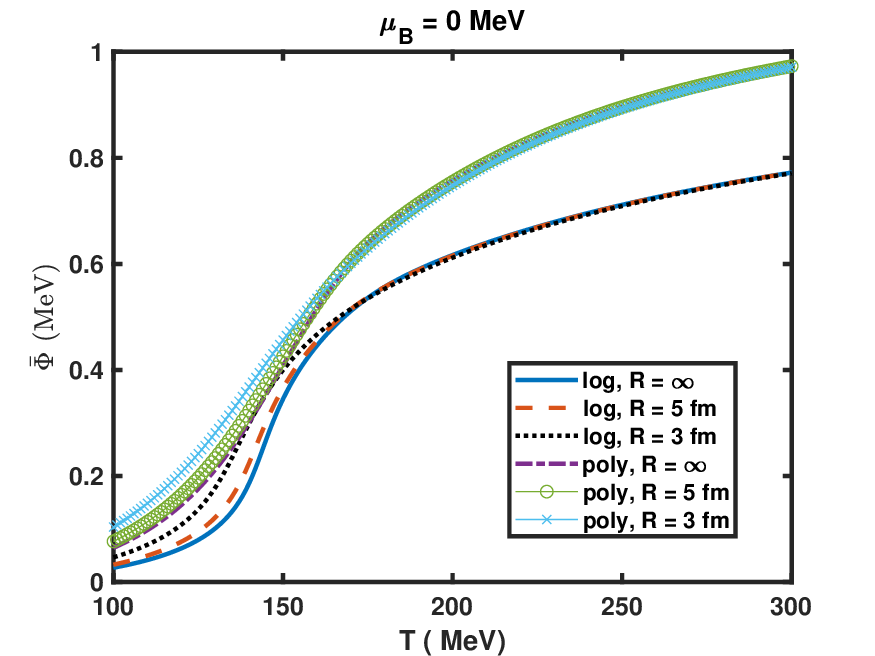}
    \hspace{0.03cm}
    (d)\includegraphics[width=7.4cm]{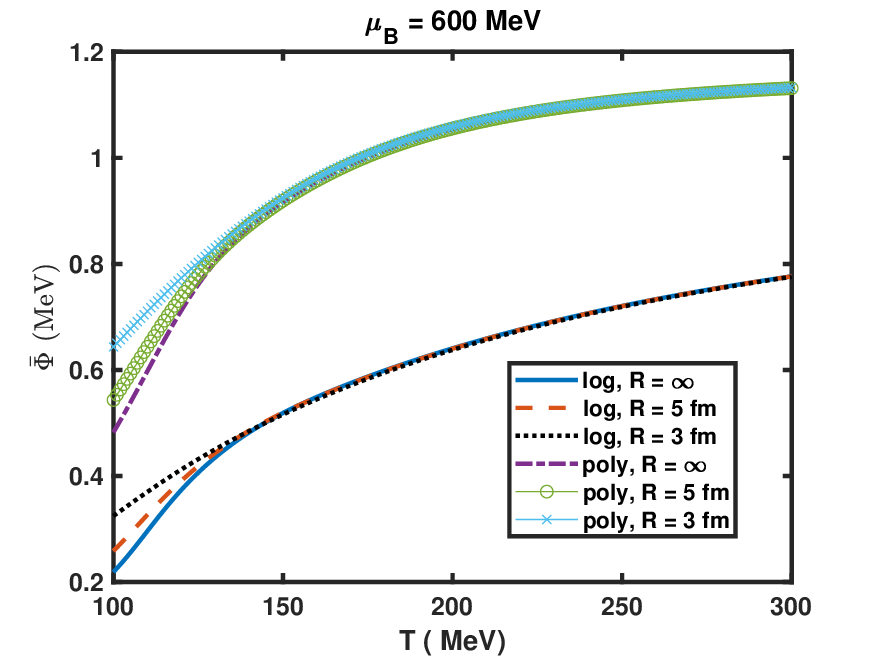}
    \hspace{0.03cm}	
    \end{minipage}
    \caption{\label{poly} \footnotesize{(Colour online) The Polyakov fields $\Phi$ and $\bar{\Phi}$ are plotted as a function of temperature $T$ for lengths of cubic volume $R=\infty$, 5 fm, and $3$ fm, at baryon chemical potential $\mu_B = 0$ MeV [in subplots (a) and (c)], and baryon chemical potential $\mu_B = 600$ MeV, isospin chemical potential $\mu_I = -30$ MeV, and strangeness chemical potential $\mu_S=125$ MeV [in subplots (b) and (d)], for both logarithmic Polyakov loop potential $\cal{U_L}$ and polynomial Polyakov loop potential $\cal{U_P}$.}}
    \end{figure*}

\begin{figure*}
    \centering
    \begin{minipage}[c]{0.98\textwidth}
    (a)\includegraphics[width=7.2cm]{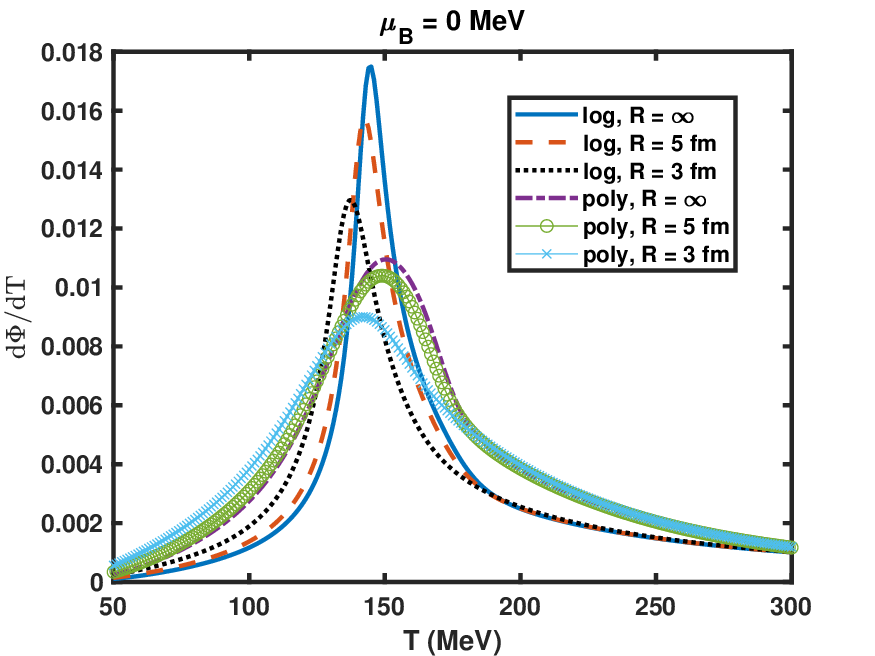}
    \hspace{0.03cm}
    (b)\includegraphics[width=7.2cm]{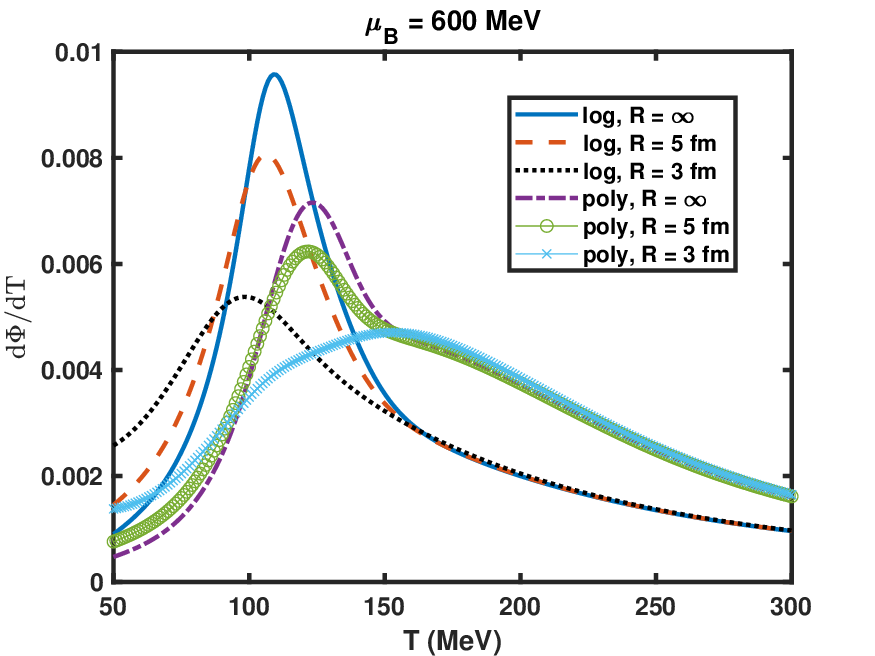}
    \hspace{0.03cm}	
    (c)\includegraphics[width=7.2cm]{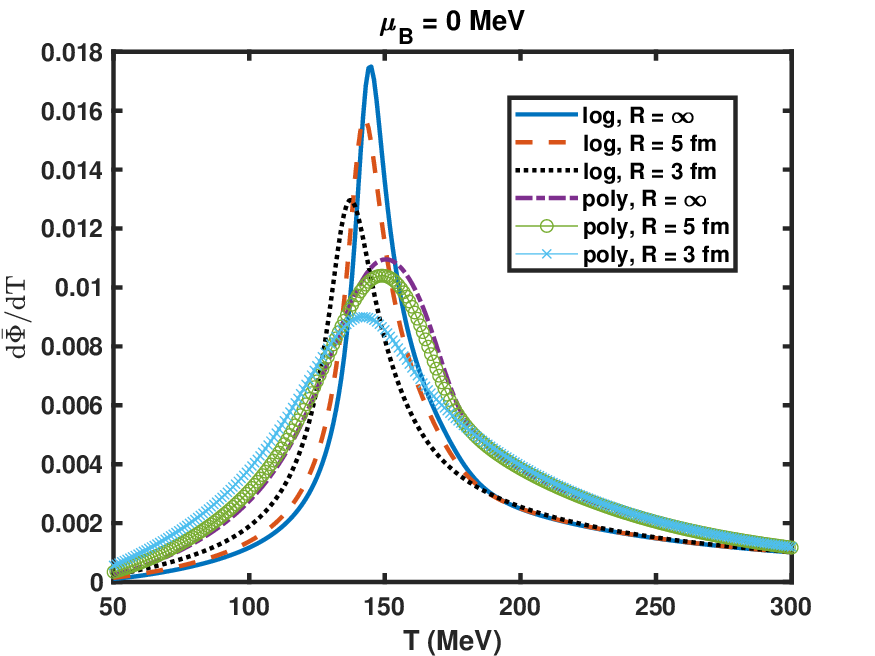}
    \hspace{0.03cm}
    (d)\includegraphics[width=7.2cm]{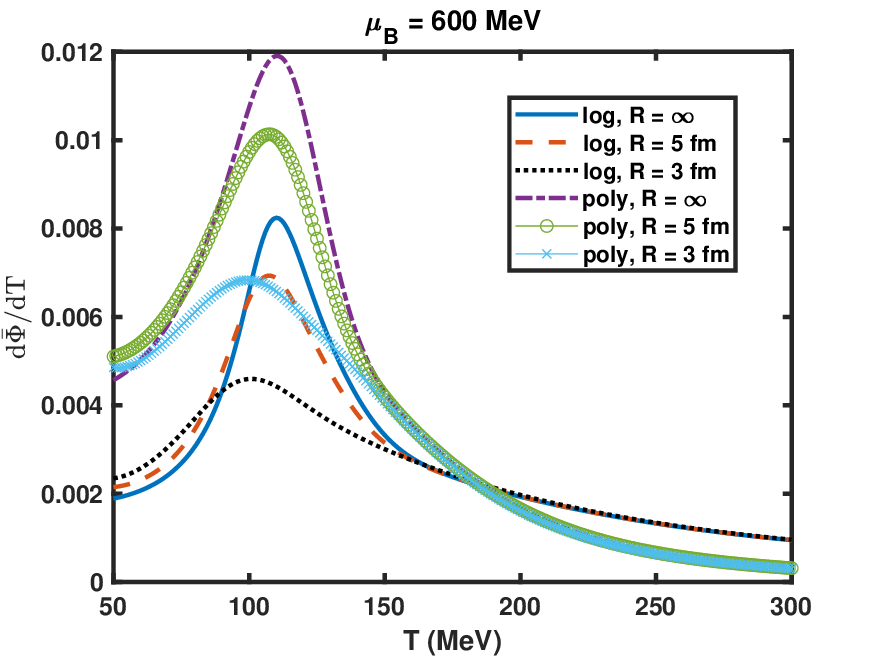}
    \hspace{0.03cm}	
    \end{minipage}
    \caption{\label{polyd} (Colour online) The derivative of Polyakov loop fields $\Phi$ and $\bar{\Phi}$ as a function of temperature $T$ with $\mu_B = 0$ MeV [in subplots (a) and (c)] and $\mu_B=600$ MeV, $\mu_I = -30$ MeV, $\mu_S=125$ MeV [in subplots (b) and (d)] at length of cubic volume $R = \infty$, 5 fm, and 3 fm, for both Polyakov loop potentials $\cal{U_L}$ and $\cal{U_P}$.}
    \end{figure*}

The Polyakov fields $\Phi$ and $\bar{\Phi}$ are plotted in Fig. \ref{poly} as a function of temperature $T$ for Polyakov loop potentials $\cal{U_L}$ and $\cal{U_P}$ at baryon chemical potentials $\mu_B = 0$ and 600 MeV, for the lengths of the cubic volume $R =\infty$, 5 fm, and 3 fm. At $\mu_B = 0$, in subplots (a) and (c) of Fig. \ref{poly}, we note that the value of $\Phi$ and $\bar{\Phi}$ is nearly zero at lower temperatures. This is expected as $\Phi (\bar{\Phi})$ serves as an order parameter of the deconfinement phase transition, and at lower temperatures, quarks remain confined within hadrons. As the temperature rises, quarks become deconfined and the magnitude of $\Phi$ and $\bar{\Phi}$ becomes non-zero. 
Unlike the scalar fields in Fig. \ref{fields}, we observe that at lower temperatures, the magnitude of $\Phi$ ($\bar{\Phi}$) rises to a small extent with the decrease in the system size. This increase in the strength of the Polyakov fields, $\Phi$ and $\bar{\Phi}$ at lower temperatures, for $\mu_B=0$ MeV, may imply that in the vicinity of the transition temperature, the tendency of the system to become deconfined is somewhat higher when the system has a finite size ($R=3$ fm) compared to a system with infinite size ($R=\infty$). This can be seen in Fig. \ref{polyd}, which shows the temperature derivatives $d\Phi/dT$ ($d\bar{\Phi}/dT$) plotted as a function of temperature $T$ for baryon chemical potentials $\mu_B = 0$ and 600 MeV, and system sizes $R = \infty$, 5 fm, and 3 fm, for Polyakov loop potentials $\cal{U_L}$ and $\cal{U_P}$. For $\mu_B=0$ MeV, in subplots (a) and (c) of Fig. \ref{polyd}, we find that the peak in $d\Phi/dT$ coincides with the peak in $d\bar{\Phi}/dT$. The position of the peak in $d\Phi/dT$ ($d\bar{\Phi}/dT$) can be used to obtain the pseudo-critical temperature for the deconfinement phase transition, $T_d$. Reducing the system size from $R=\infty$ to 5 fm to 3 fm leads to a drop in the pseudo-critical deconfinement temperature $T_d$ from $145$ (150) MeV to $143$ (149) MeV to 137 (142) MeV, for $\cal{U_L}$ ($\cal{U_P}$). That is, the deconfinement transition temperature $T_d$ may be shifted towards lower $T$ at lower $R$. However, as is evident in Fig. \ref{poly}, at higher values of $T$, the magnitude of $\Phi$ $(\bar{\Phi})$ becomes independent of the system size. This may imply that the deconfinement phase transition becomes independent of the size of the system at very high temperatures. 

In the case of finite baryon chemical potential, $\mu_B=600$ MeV (Figs. \ref{poly}(b) and \ref{poly}(d)), the magnitude of the Polyakov fields, $\Phi$ and $\bar{\Phi}$ are found to be non-zero even at lower temperatures. This may indicate a shift in the deconfinement temperature to lower temperature values for systems with non-vanishing baryon chemical potential.
From the subplots (b) and (d) of Fig. \ref{polyd}, we find that for infinite system size ($R=\infty$), increasing the baryon chemical potential from 0 to 600 MeV causes the pseudo-critical deconfinement temperature $T_d$ to change to much lower temperature values ($T_d = 109$ (123) MeV, for $\cal{U_L}$ ($\cal{U_P}$)).
As for the impact of the system size on the Polyakov fields $\Phi$ and $\bar{\Phi}$, we see that the peak in $d\Phi/dT$ ($d\bar{\Phi}/dT$) is shifted to much lower values at $R=3$ fm. This may imply that the deconfinement temperature $T_d$ is reduced for system with finite size and baryon chemical potential. 

\begin{figure*}
    \centering
    \begin{minipage}[c]{0.98\textwidth}
    (a)\includegraphics[width=7.2cm]{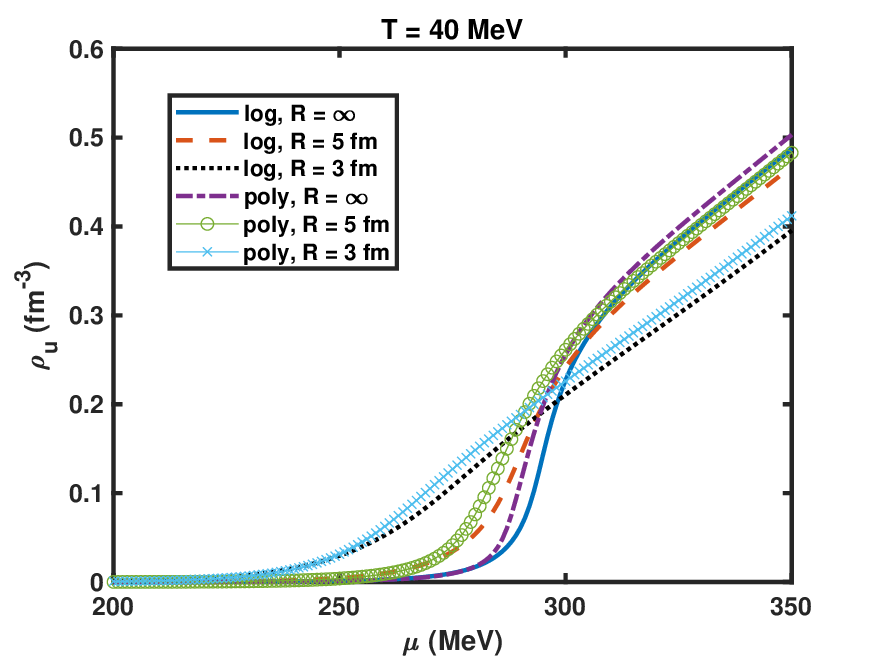}
    \hspace{0.03cm}
    (b)\includegraphics[width=7.2cm]{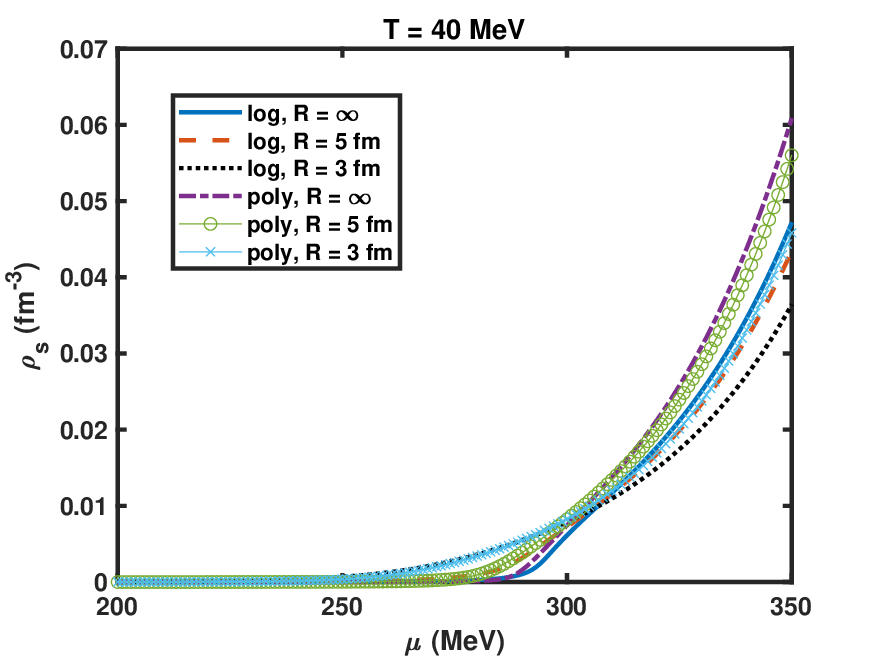}
    \hspace{0.03cm}	
    (c)\includegraphics[width=7.2cm]{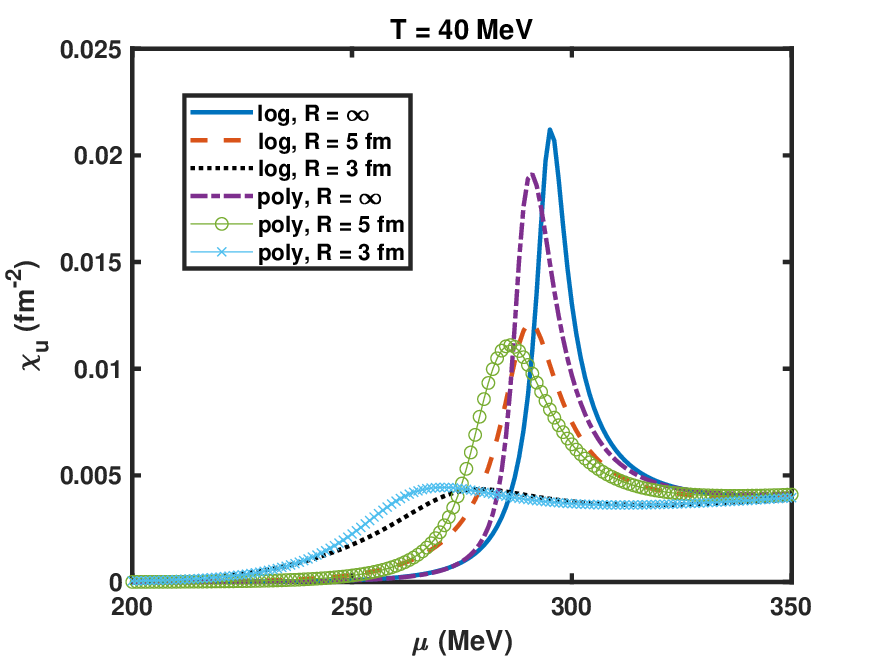}
    \hspace{0.03cm}
    (d)\includegraphics[width=7.2cm]{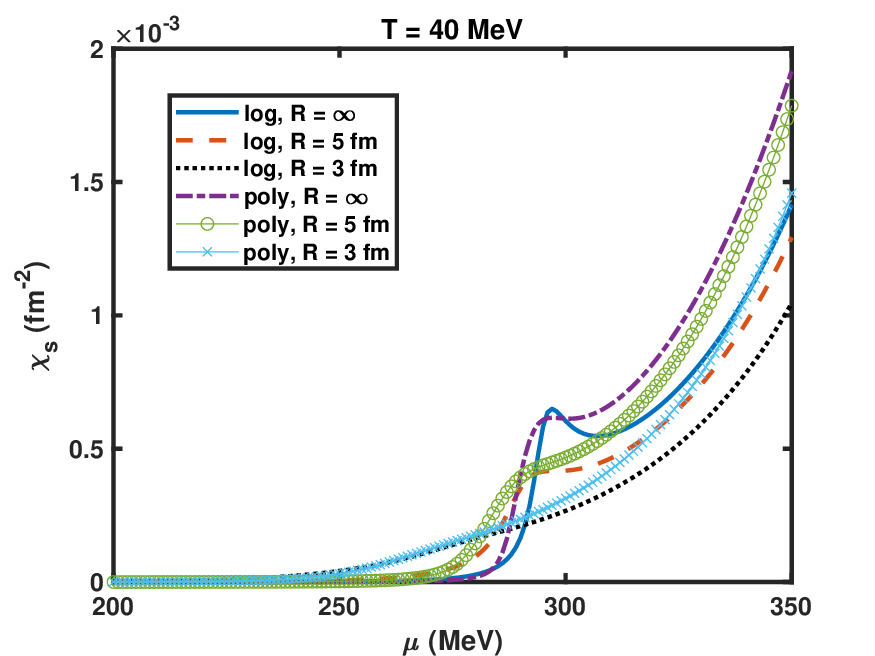}
    \hspace{0.03cm}	
    \end{minipage}
    \caption{\label{figrho} (Colour online) The number density of the up quark, $\rho_u$ and its susceptibility, $\chi_u$ [in subplots (a) and (c)] and the number density of the strange quark, $\rho_s$ and its susceptibility, $\chi_s$ [in subplots (b) and (d)] are plotted as a function of quark chemical potential $\mu$ at $T = 40$ MeV and system sizes $R = \infty$, 5 fm, and 3 fm, for both Polyakov loop potentials $\cal{U_L}$ and $\cal{U_P}$.}
    \end{figure*} 

\begin{figure*}
\centering
\begin{minipage}[c]{0.98\textwidth}
(a)\includegraphics[width=7.2cm]{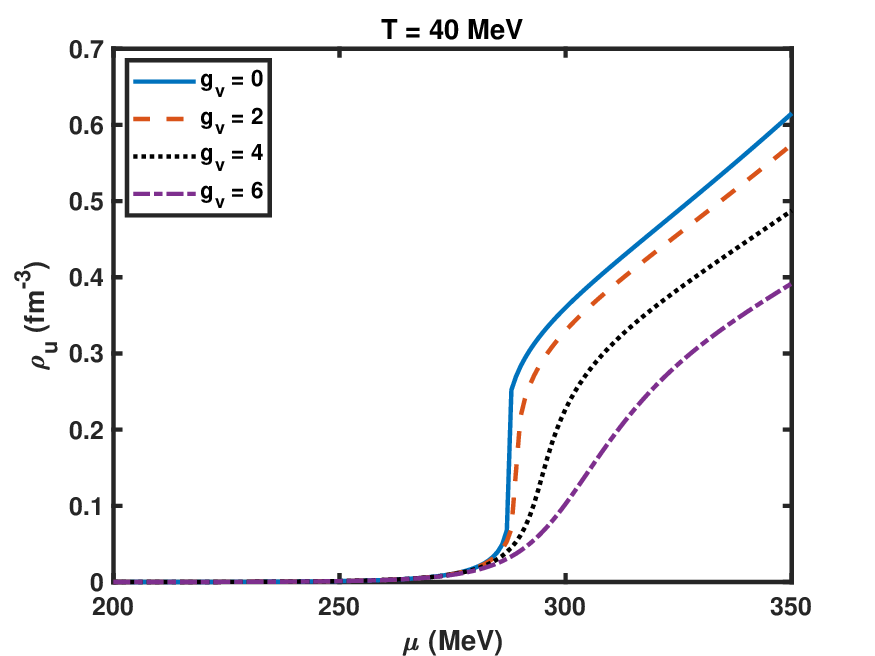}
\hspace{0.03cm}
(b)\includegraphics[width=7.2cm]{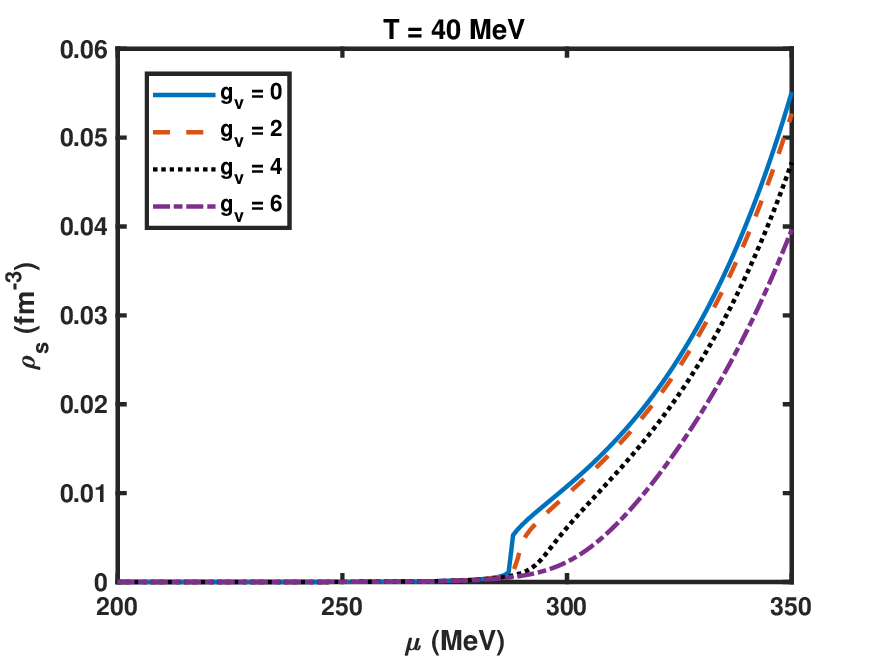}
\hspace{0.03cm}	
(c)\includegraphics[width=7.2cm]{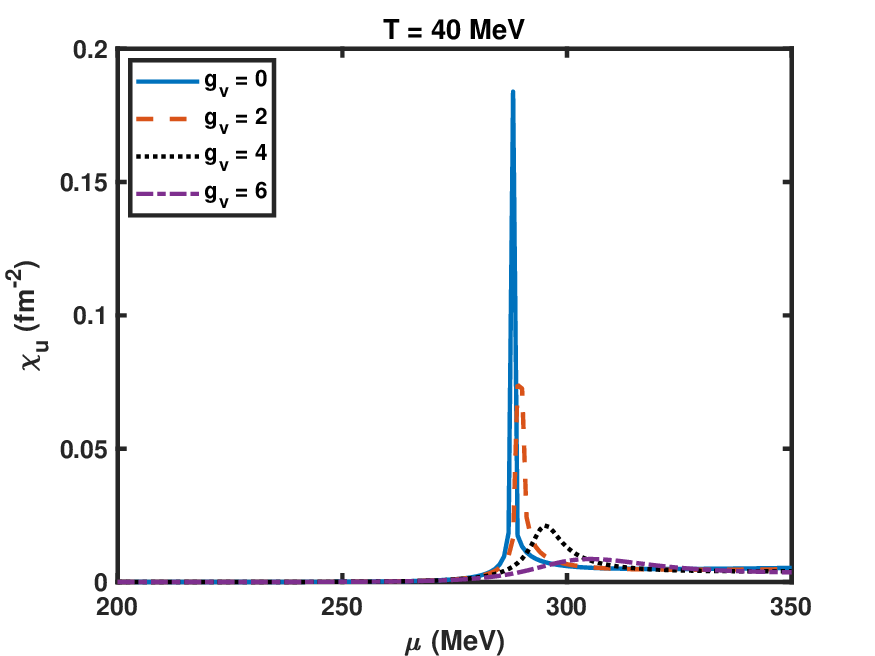}
\hspace{0.03cm}
(d)\includegraphics[width=7.2cm]{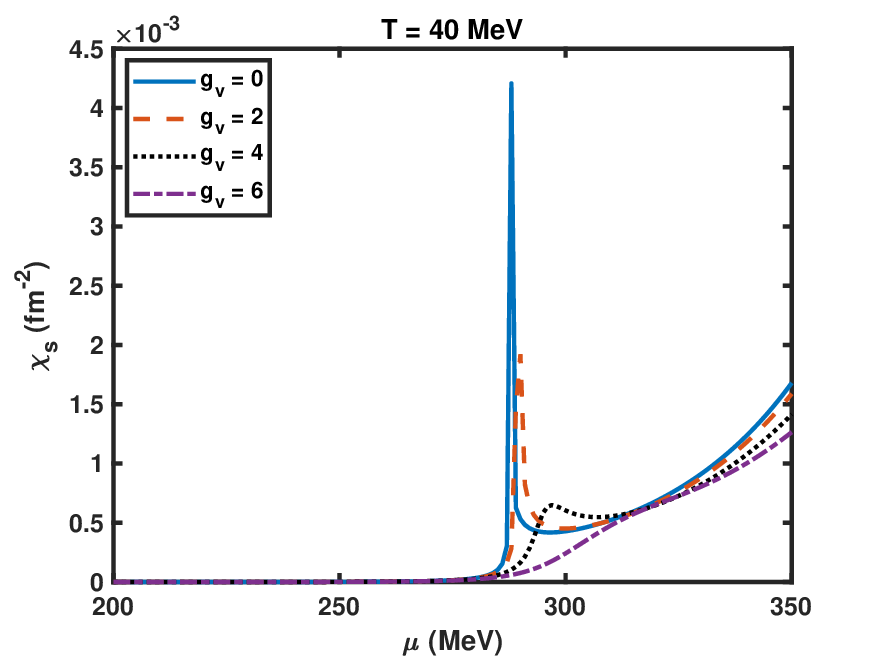}
\hspace{0.03cm}	
\end{minipage}
\caption{\label{figrhod} (Colour online) The number density of the up quark, $\rho_u$ and its susceptibility, $\chi_u$ [in subplots (a) and (c)] and the number density of the strange quark, $\rho_s$ and its susceptibility, $\chi_s$ [in subplots (b) and (d)] are plotted as a function of quark chemical potential $\mu$ at vector coupling strength $g_v=0,$ 2, 4, and 6, for fixed $T = 40$ MeV, $R = \infty$, and Polyakov loop $\cal{U_L}$.}
\end{figure*}

In subplots (a) and (c) of Fig. \ref{figrho}, we have plotted the quark number density of the $u$ quark, $\rho_u$ and its susceptibility, $\chi_u=\frac{\partial \rho_u}{\partial \mu}$, as a function of the quark chemical potential $\mu$, for system sizes $R = \infty$, 5 fm, and 3 fm, and Polyakov loop potentials $\cal{U_L}$ and $\cal{U_P}$ at fixed temperature, $T = 40$ MeV. The discontinuity in quark number density $\rho_u$ can serve as an order parameter to study the first order chiral phase transition \cite{mao}. We can see that $\rho_u$ remains negligible at low $\mu_B$ and shows a sudden rise at higher quark chemical potential. The rise in $\rho_u$ leads to a peak in $\chi_u$, which can be used as an indicator of the first order phase transition. 
We find that for $T = 40$ MeV and $R = \infty$, $\chi_u$ shows divergence at $\mu = 291$ MeV for Polyakov loop $\cal{U_P}$, indicating a first order phase transition. However, for the logarithmic Polyakov loop potential $\cal{U_L}$, we observe a peak in $\chi_u$ at $\mu=295$ MeV. As we decrease $R$ to 5 fm, we find that the peak in $\chi_u$ is shifted to $\mu = $ 286 (291) MeV for $\cal{U_P}$ ($\cal{U_L}$). Furthermore, we find no discontinuity in $\rho_u$, or corresponding divergence in $\chi_u$, for either Polyakov loop. The peak in $\chi_u$ completely vanishes as we further decrease the system size to $R=3$ fm. This may imply a change in the order of the phase transition at $R=3$ fm. The change from first order phase transition to a crossover can also be understood from Fig. \ref{figrhod}, which shows the effect of vector coupling $g_v$ on the order of the phase transition at infinite system size ($R=\infty$) and logarithmic Polyakov loop potential $\cal{U_L}$. At temperature $T=40$ MeV and $g_v = 0$, we observe a divergence in $\chi_u$ at $\mu = 288$ MeV. Increasing the vector coupling to $g_v = 2$ and 4 shifts the peak in $\chi_u$ to $\mu = 289$ MeV and 295 MeV, respectively. 
However, as the value of $g_v$ is increased to $g_v = 6$, we find that the divergence in $\chi_u$ disappears. This may indicate that the first order chiral phase transition of the $u$ quark vanishes and remains a crossover for higher $g_v$ even at higher chemical potential. 
In Ref. \cite{friesen}, the PNJL model was used to study the influence of vector interaction on the critical end point (CEP). They reported that increasing the vector interaction strength causes the CEP to move to lower temperatures and higher chemical potentials.

 The subplots (b) and (d) of Fig. \ref{figrho} show the quark number density, $\rho_s$, of the $s$ quark and its susceptibility, $\chi_s = \frac{\partial \rho_s}{\partial \mu}$, as a function of quark chemical potential $\mu$ for $R = \infty$, 5 fm, and 3 fm, and $\cal{U_L}$
  at temperature $T = 40$ MeV. We observe that for $R = \infty$, $\rho_s$ remains negligible and starts rising at higher values of quark chemical potential. However, we do not observe any discontinuity in $\rho_s$ or divergence in $\chi_s$, implying that the chiral phase transition of the $s$ quark remains a crossover. This is similar to the findings in Refs. \cite{tawfik,costa2019}, where the chiral phase transition of the $s$ quark remains a crossover over the entire QCD phase diagram. However, from Fig. \ref{figrhod} (b) and (d), we can see that for $T=40$ MeV and $R=\infty$, the $s$ quark experiences a first order phase transition at $\mu = $ 288 MeV and 290 MeV for $g_v = 0$ and 2, respectively. Increasing $g_v$ further results in the change of the order of phase transition from first order to a crossover. This signifies the importance of the strength of vector interaction on the order of the phase transition. Hence, we conclude that decreasing the system size results in shifting the first order phase transition of the $u$ quark to lower quark chemical potential. The order of the phase transition becomes crossover over the entire $(T,\mu_B$) range for sufficiently small systems ($R \approx$ 3 fm). Additionally, the $s$ quark shows a first order phase transition only for low values of vector coupling $g_v$. The order of phase transition of the $u$ and $s$ quark changes to crossover for higher vector coupling.

\par
\begin{figure*}
    \centering
    \begin{minipage}[c]{0.98\textwidth}
    (a)\includegraphics[width=6.4cm]{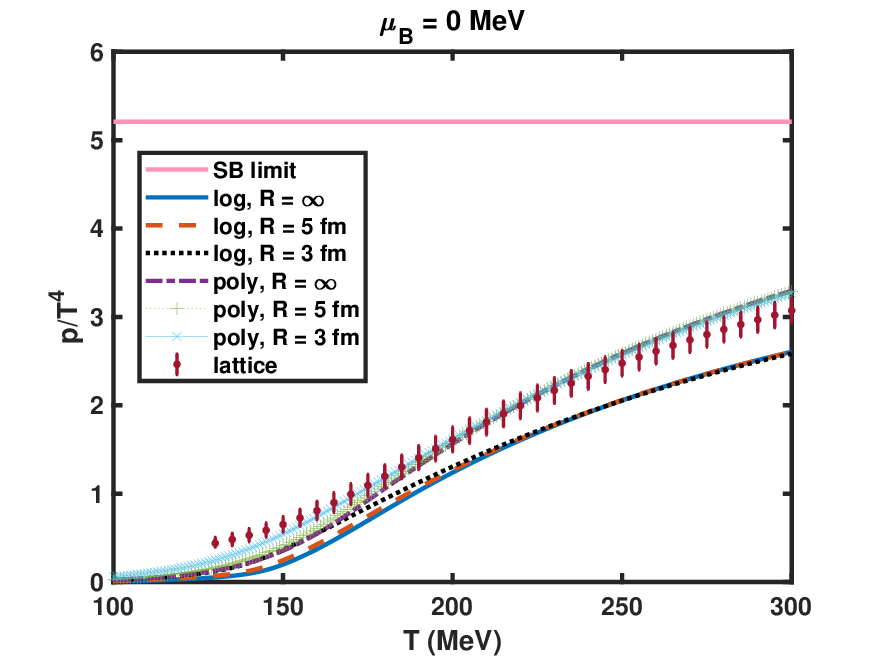}
    \hspace{0.03cm}
    (b)\includegraphics[width=6.4cm]{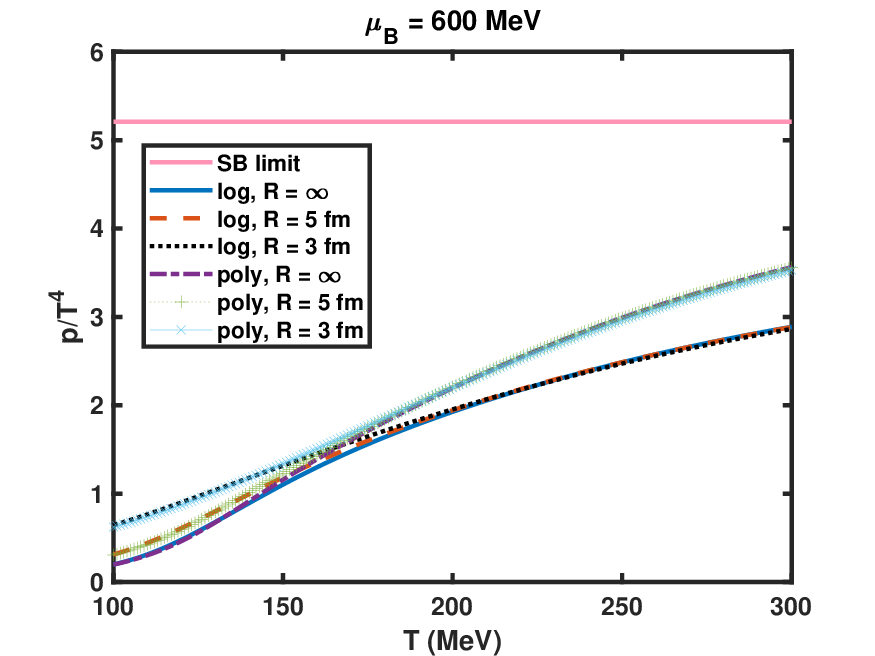}
    \hspace{0.03cm}	
    (c)\includegraphics[width=6.4cm]{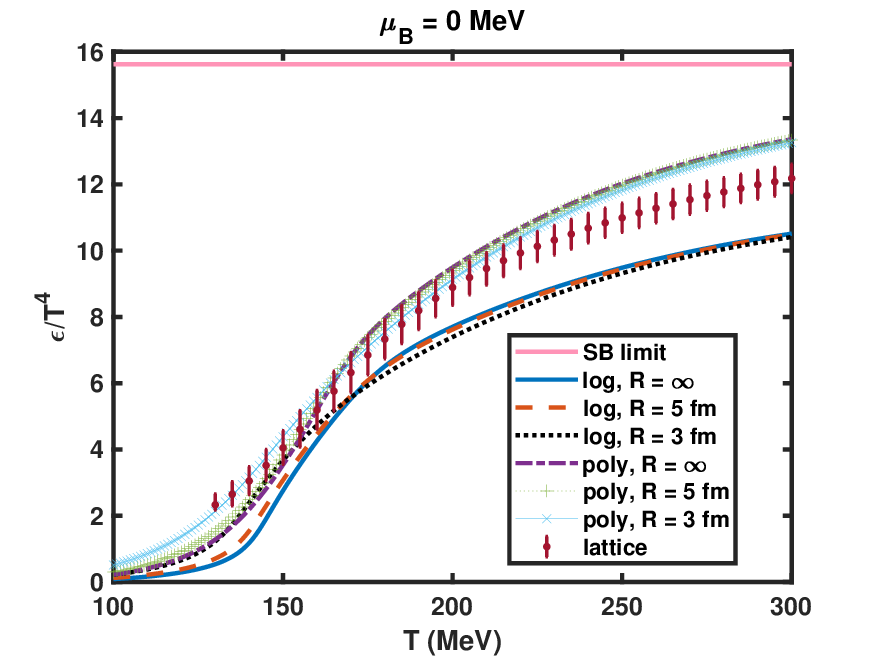}
    \hspace{0.03cm}
    (d)\includegraphics[width=6.4cm]{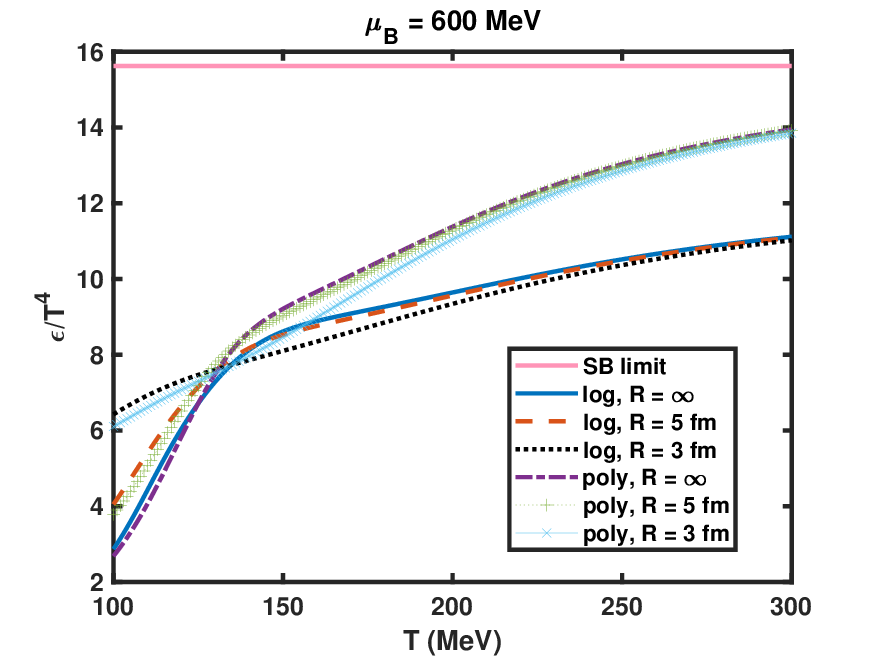}
    \hspace{0.03cm}	
    (e)\includegraphics[width=6.4cm]{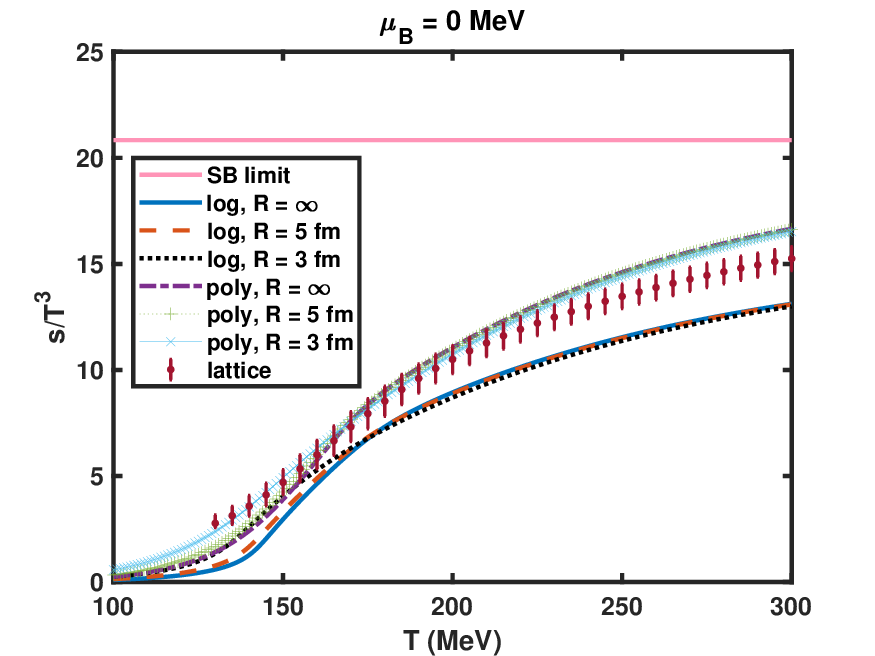}
    \hspace{0.03cm}
    (f)\includegraphics[width=6.4cm]{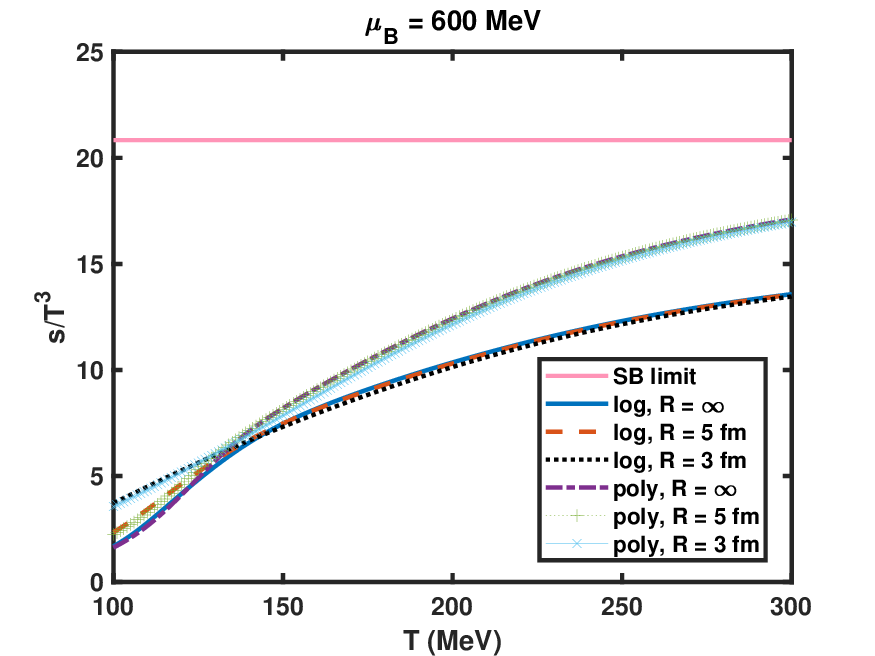}
    \hspace{0.03cm}
    (g)\includegraphics[width=6.4cm]{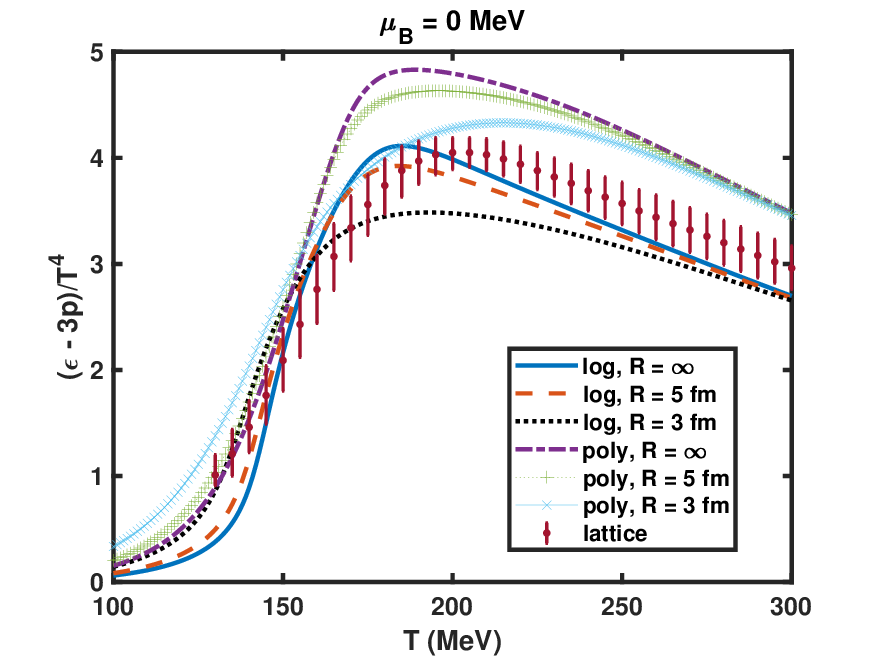}
    \hspace{0.03cm}
    (h)\includegraphics[width=6.4cm]{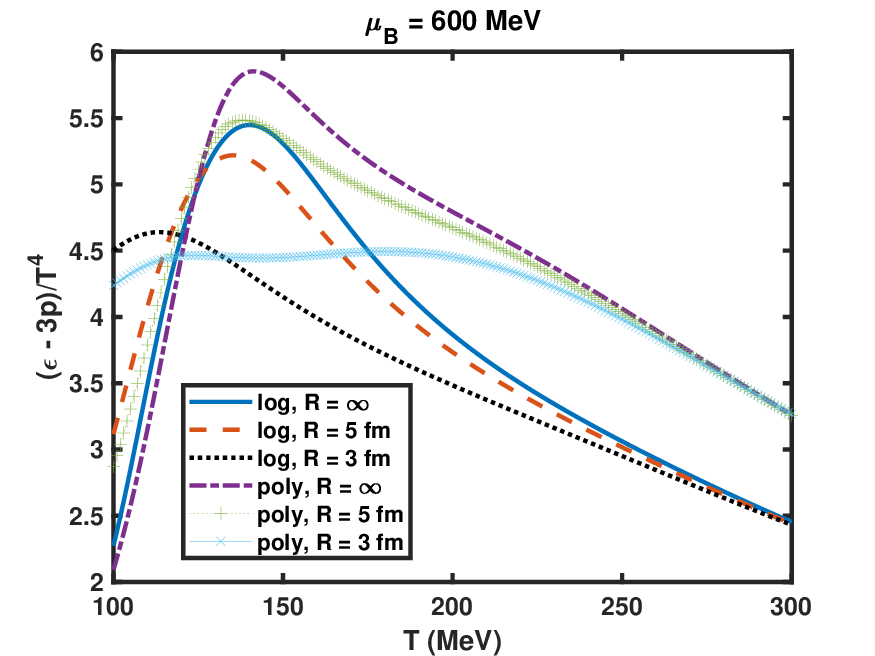}
    \hspace{0.03cm}
    \end{minipage}
    \caption{\label{thermo} \scriptsize{(Colour online) The scaled pressure $p/T^4$, scaled energy density $\epsilon/T^4$, scaled entropy density $s/T^3$, and trace anomaly $(\epsilon-3p)/T^4$ as a function of temperature $T$ for different length of cubic volume $R=\infty$, 5 fm, and 3 fm, at $\mu_B = 0$ MeV [in subplots (a), (c), (e), and (g)] and $\mu_B=600$ MeV, $\mu_I = -30$ MeV, $\mu_S=125$ MeV [in subplots (b), (d), (f), and (h)], for Polyakov loop potentials $\cal{U_L}$ and $\cal{U_P}$.}}
    \end{figure*}

 In Fig. \ref{thermo}, we have plotted the variations of scaled thermodynamic quantities: pressure $p/T^4$, energy density $\epsilon/T^4$, entropy density $s/T^3$, and trace anomaly $(\epsilon-3p)/T^4$ as a function of temperature $T$ for system sizes $R=\infty$, 5 fm, and 3 fm, at baryon chemical potentials $\mu_B=0$ and 600 MeV, with Polyakov loops $\cal{U_L}$ and $\cal{U_P}$. The results are compared to the lattice data \cite{baza2009,baza2014} for $\mu_B = 0$ MeV.
All these quantities remain negligible in the low-temperature regime but suddenly rise as the system approaches the transition temperature. For high $T$, the interaction between quarks becomes weaker, and the thermodynamic quantities approach the ideal gas limit or the Stefann-Boltzmann (SB) limit, which depends on the number of flavours considered \cite{borsanyi,manisha}.
The magnitude of all the thermodynamic quantities is found to be higher for the polynomial Polyakov loop potential $\cal{U_P}$ than for the logarithmic Polyakov loop potential $\cal{U_L}$. Crucially, they all remain within the SB limit for both Polyakov loops $\cal{U_L}$ and $\cal{U_P}$ at high temperatures. All the thermodynamic quantities are found to be in good qualitative agreement with the lattice data.
Notably, decreasing the system size from $R=\infty$ to $R=3$ fm leads to an increase in the magnitude of the thermodynamic quantities below the transition temperature $T_c$. A similar impact of finite size is observed on the thermodynamic quantities (increasing value for smaller size systems) using the PNJL model \cite{saha2018, grunfeld}. However, it contrasts with the results previously obtained in the PCQMF model, where decreasing the system size leads to a decrease in the thermodynamic quantities \cite{nisha}. This is the consequence of including the fermionic vacuum term in the current study. 
We should point out here that going beyond the transition temperature leads to a decrease in the thermodynamic quantities for smaller systems. However, at very high temperatures, the effect of finite size almost vanishes.
Increasing baryon chemical potential $\mu_B$ to 600 MeV, in subplots (b), (d), (f), and (h) of Fig. \ref{thermo}, we observe that the magnitude of the thermodynamic quantities becomes non-zero even at lower temperatures. However, they all remain within the SB limit at higher $T$. 
Again, this may indicate that the transition temperature shifts to lower temperature values for higher baryon chemical potentials. Reducing the system size from $R=\infty$ to $R = 3$ fm, we note that the magnitude of these quantities is enhanced, just like in the case of vanishing baryon chemical potential.
So, we conclude that there is a slight enhancement in the values of thermodynamic quantities as the system size decreases, which is more significant for temperatures near or below the transition temperature.


\begin{figure*}
    \centering
    \begin{minipage}[c]{0.98\textwidth}
    (a)\includegraphics[width=7.4cm]{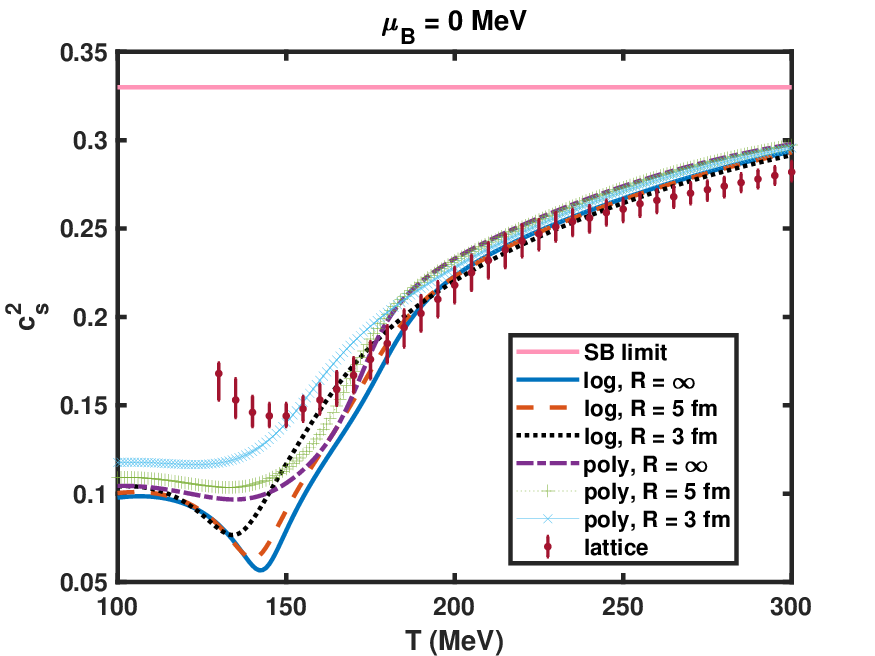}
    \hspace{0.03cm}
    (b)\includegraphics[width=7.4cm]{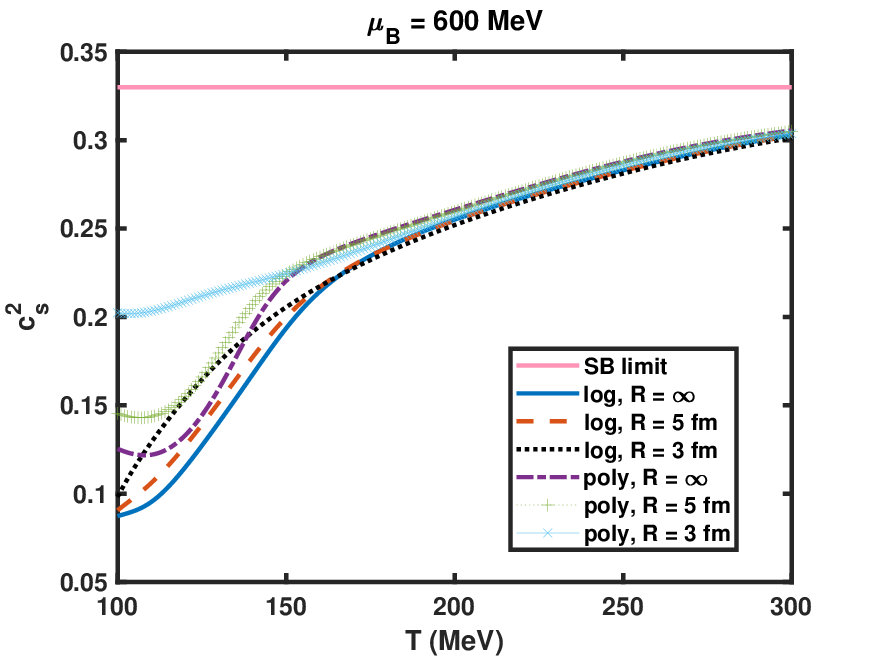}
    \hspace{0.03cm}	
    (c)\includegraphics[width=7.4cm]{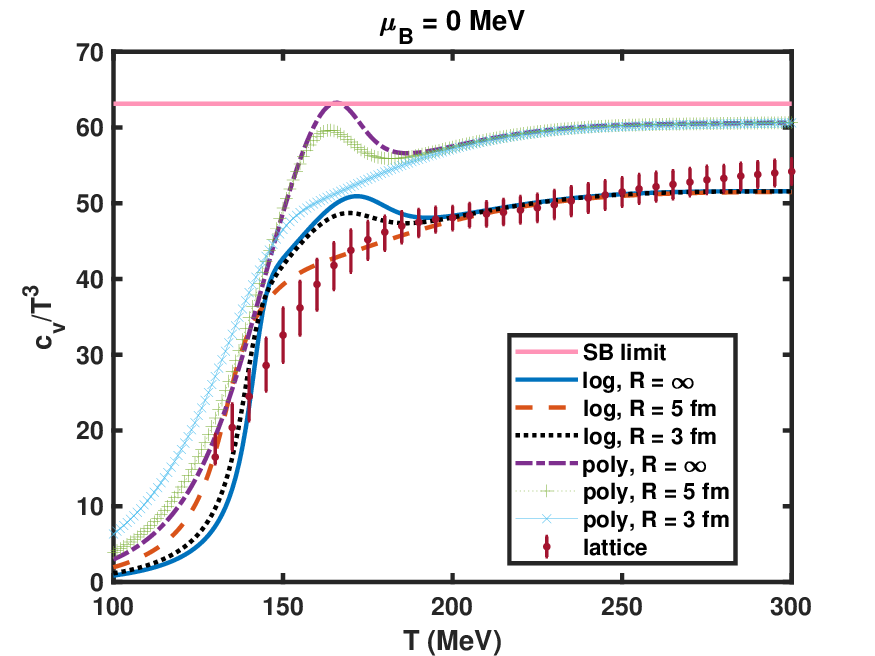}
    \hspace{0.03cm}
    (d)\includegraphics[width=7.4cm]{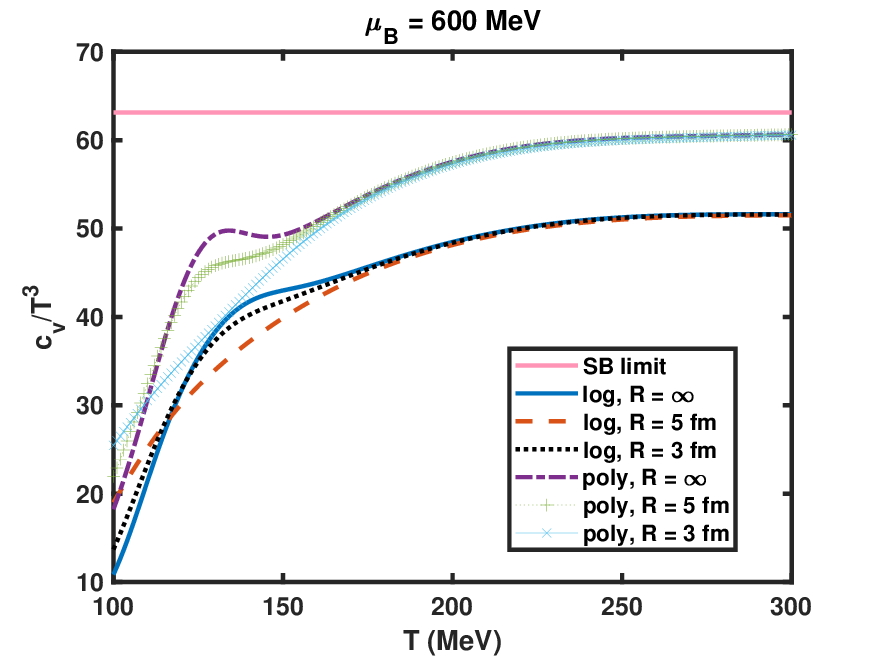}
    \hspace{0.03cm}	
    \end{minipage}
    \caption{\label{sound} (Colour online) The speed of sound squared $c_s^2$ and scaled specific heat $c_v/T^3$ as a function of temperature $T$ for different lengths of cubic volume $R=\infty$, 5 fm, and 3 fm, at $\mu_B = 0$ MeV [in subplots (a) and (c)] and $\mu_B = 600$ MeV, $\mu_I = -30$ MeV, and $\mu_S=125$ MeV [in subplots (b) and (d)], for Polyakov loop potentials $\cal{U_L}$ and $\cal{U_P}$.}
    \end{figure*}

Fig. \ref{sound} showcases the temperature dependence of the square of the speed of sound $c_s^2$ and scaled specific heat $c_v/T^3$, for system sizes $R = \infty$, 5 fm, and 3 fm, at zero baryon chemical potential $\mu_B=0$ MeV (subplots (a) and (c)) and finite baryon chemical potential $\mu_B = 600$ MeV (subplots (b) and (d)), for Polyakov loops $\cal{U_L}$ and $\cal{U_P}$. For $\mu_B = 0$ MeV, we observe that the square of the speed of sound $c_s^2$ shows good qualitative agreement with the lattice data, reaching a minimum near the transition temperature and then rising to approach its SB limit $c_s^2 = 1/3$ at high temperatures. We observe that the value of $c^2_s$ is increased for smaller systems ($R<\infty$). Additionally, the position of the minimum is shifted towards a lower temperature value as the system's size becomes smaller.
As we decrease the system size from $R=\infty$ to 5 fm and 3 fm, we observe that the minimum in $c_s^2$ is shifted from 142 MeV to 140 MeV and 134 MeV, respectively for the logarithmic Polyakov loop potential $\cal{U_L}$. A similar effect of finite size is observed for the speed of sound in the PNJL model \cite{saha2018}.
Notably, the dip in the value of $c_s^2 $ becomes less distinguished for the polynomial form of the Polyakov loop potential $\cal{U_P}$. In subplot (b) of Fig. \ref{sound}, for the finite baryon chemical potential $\mu_B=600$ MeV, we find that the minimum in $c_s^2$ is reached at much lower values of temperatures, indicating a shift in the transition temperature to lower $T$, at higher $\mu_B$. 

Next, we discuss the impact of finite size on the temperature dependence of scaled specific heat $c_v/T^3$ at baryon chemical potential $\mu_B = 0$ MeV for Polyakov loops $\cal{U_L}$ and $\cal{U_P}$ in subplot (c) of Fig. \ref{sound}. The specific heat plays a crucial role in thermodynamics, as it can be viewed as a measure of how a system responds to phase transition \cite{zhang}.
For the logarithmic form of the Polyakov loop potential $\cal{U_L}$, we observe that the value of $c_v/T^3$ matches very well with the lattice data and shows a small fluctuation near the transition temperature and then rises slowly before reaching its SB limit (63.15) at high temperatures. For the polynomial Polyakov loop $\cal{U_P}$, the specific heat is found to be larger in magnitude (but still within the SB limit at high $T$) and in qualitative agreement with the lattice data.
Interestingly, including the quark back reaction in the PCQMF model results in the disappearance of the two peak structures usually observed in $c_v/T^3$ \cite{zhang,grunfeld,zhao,dj}.     
The magnitude of $c_v/T^3$ increases, and the position of the fluctuation is shifted to lower temperatures when we decrease the system size.
The PNJL model shows a similar effect of finite size on the specific heat \cite{grunfeld}. Concretely, the nature of the specific heat becomes smoother as we decrease the system size to $R=3$ fm.
For $\cal{U_P}$, the value of $c_v/T^3$ almost reaches the SB limit before returning to it at high $T$.
Notably, the value of $c_v/T^3$ becomes independent of the system size at high $T$. In Fig. \ref{sound}(d), increasing the baryon chemical potential $\mu_B$ to 600 MeV, we observe that $c_v/T^3$ becomes non-zero at low $T$ and shows a smooth rise with increasing temperature before saturating to a constant value (less than its SB limit) at higher $T$. 


\par
\begin{figure*}
    \centering
    \begin{minipage}[c]{0.98\textwidth}
    (a)\includegraphics[width=7.4cm]{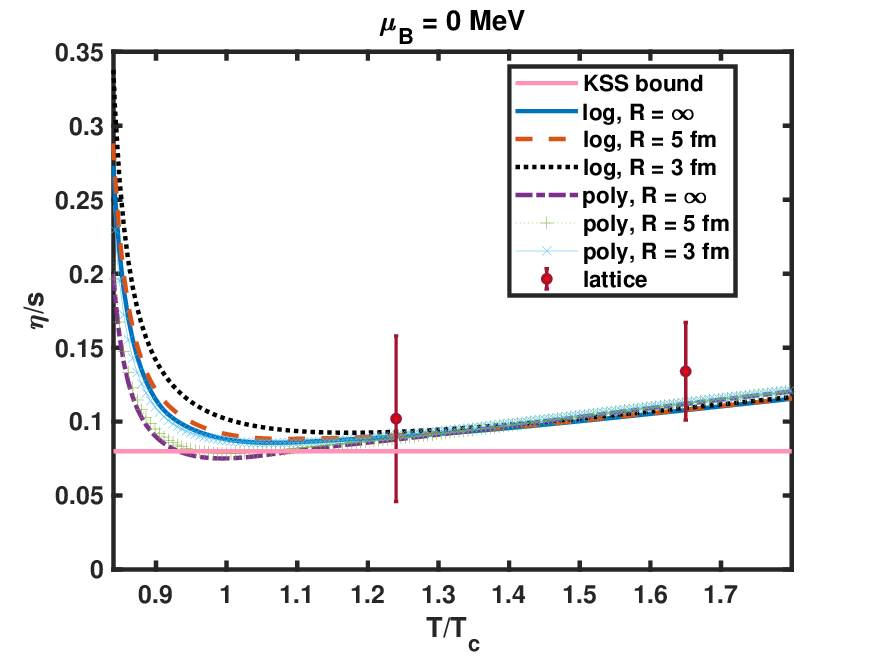}
    \hspace{0.03cm}
    (b)\includegraphics[width=7.4cm]{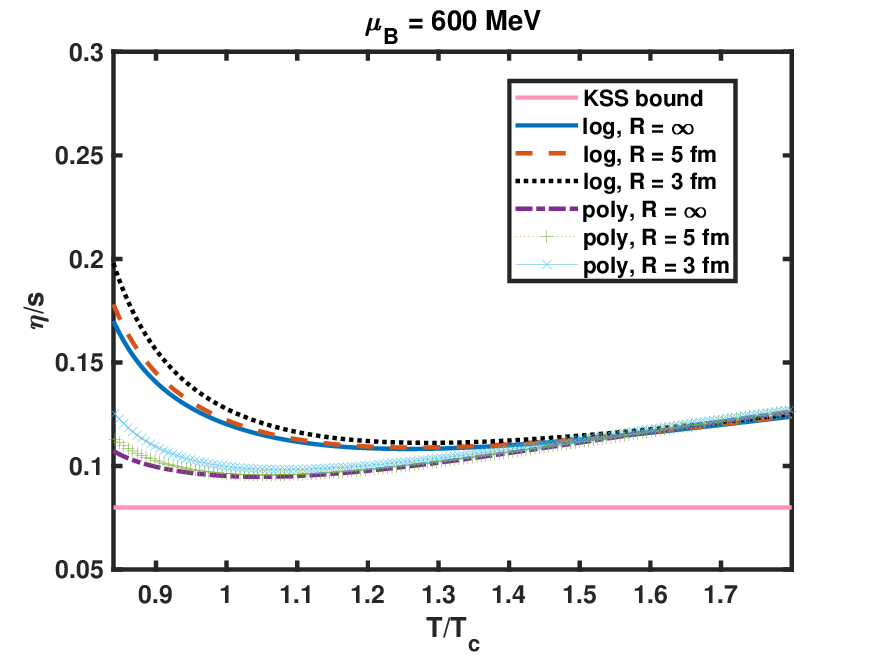}
    \hspace{0.03cm}	
    (c)\includegraphics[width=7.4cm]{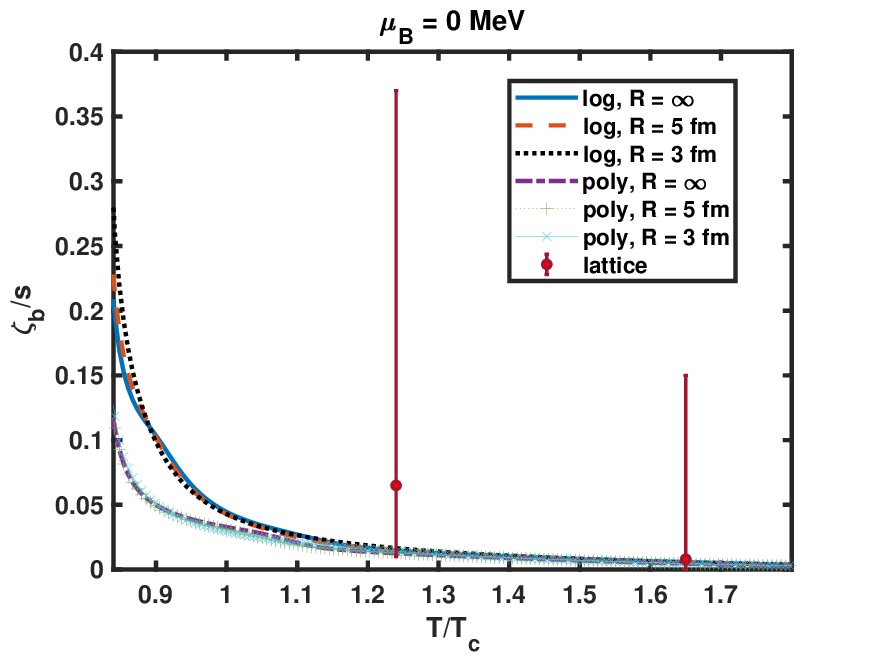}
    \hspace{0.03cm}
    (d)\includegraphics[width=7.4cm]{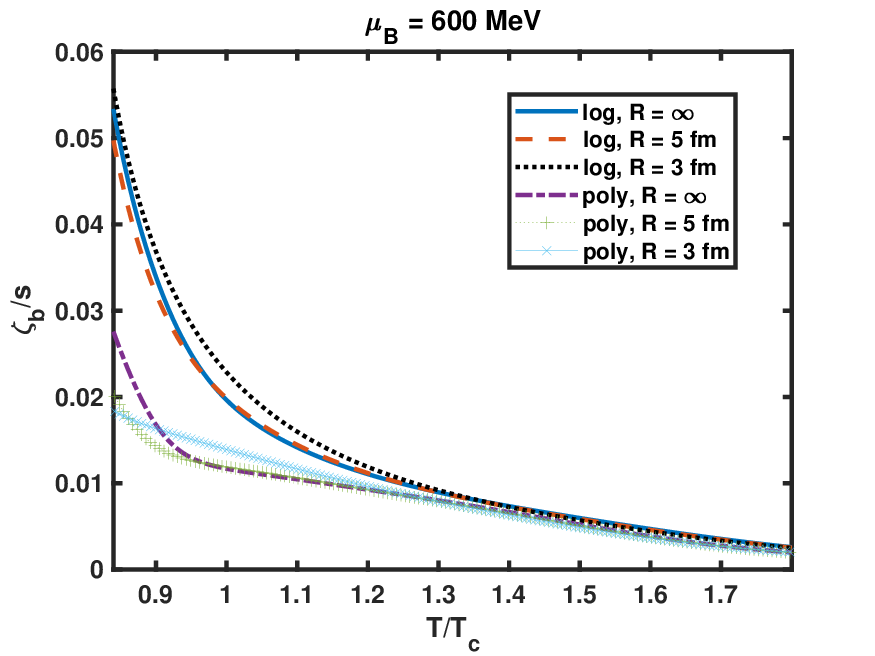}
    \hspace{0.03cm}	
    \end{minipage}
    \caption{\label{vis} (Colour online) The specific shear viscosity $\eta/s$ and normalised bulk viscosity $\zeta_b/s$ as a function of $T/T_c$ for system sizes $R=\infty$, 5 fm, and 3 fm, at $\mu_B = 0$ MeV compared to lattice data from Refs. \cite{meyer2007,meyer2008} [in subplots (a) and (c)] and $\mu_B = 600$ MeV, $\mu_I = -30$ MeV, and $\mu_S=125$ MeV [in subplots (b) and (d)], for Polyakov loop potentials $\cal{U_P}$ and $\cal{U_L}$.}
    \end{figure*}

 Now, we begin the discussion of the viscous properties of the QCD matter, namely the shear viscosity $\eta$ and the bulk viscosity $\zeta_b$. These properties are closely related to the hydrodynamic flow of the relativistic fluid (QGP) and the lateral movement of its constituents (partons) as the system undergoes expansion. In hydrodynamics, $\eta$ acts as a quantitative measure of a fluid's ability to resist shear deformation \cite{wondrak}. Changes in $\eta$ and $\zeta_b$ near the transition temperature $T_c$ modify the hydrodynamic evolution of the medium and thus influence physical observables such as elliptic flow and correlation functions \cite{roma,luzum,olli}. Fig. \ref{vis} shows the variations of the specific shear viscosity $\eta/s$ and normalised bulk viscosity $\zeta_b/s$ with $T/T_c$ for different system sizes ($R = \infty$, 5 fm, and 3 fm) and Polyakov loops ($\cal{U_L}$ and $\cal{U_P}$), at baryon chemical potentials $\mu_B=0$ and 600 MeV. The results are compared with lattice data points from Refs. \cite{meyer2007,meyer2008}. The KSS bound, which sets a theoretical lower limit of $\eta/s$ to around $1/4\pi$ for QCD matter, is also shown \cite{kss}. 
In the PCQMF model, there are two types of phase transition: the chiral phase transition (characterised by $T_{\chi}$) and the deconfinement phase transition (characterised by $T_d$). In the present study, we define the transition temperature $T_c$ as the average of these two temperatures as in Ref. \cite{bhatt2013}. Note that different models have different values of the transition temperature, for instance, $T_c$ = 240 and 200 MeV, for the PLSM and NJL model, respectively \cite{tawfik}.
We observe that at $\mu_B=0$ MeV, the $\eta/s$ is high at lower temperatures and decreases sharply to reach a minimum around the transition temperature. We note that for $R=\infty$, the value of $\eta/s$ goes below the KSS bound near the transition temperature for the Polyakov loop $\cal{U_P}$.
Although $\eta$ (not shown here) is extremely small for temperatures below the transition temperature, $\eta/s$ diverges at lower $T$. This is similar to the results obtained for the PNJL model in Refs. \cite{saha2018,islam,ghosh2015}. In Refs. \cite{prakash,lang}, the authors show that $\eta/s$ is extremely large at lower temperatures and should diverge as $T\rightarrow0$. For higher temperatures, $\eta/s$ increases slowly and resembles a fluid undergoing a liquid-gas phase transition, which has a minimum near the transition temperature \cite{schaefer2009}. As for the finite size dependence of $\eta/s$, we found that specific shear viscosity $\eta/s$ gets enhanced, and the dip faint for smaller system sizes ($R=5$ and $3$ fm). 
Our results suggest a similar phenomenon to that in Refs. \cite{roy,saha2018}, where the value of $\eta/s$ is enhanced for lower system sizes compared to a system with infinite size.
For a finite value of chemical potential $\mu_B=600$ MeV, the qualitative behaviour of $\eta/s$ remains the same as in the $\mu_B = 0$ case. We also note that the value of $\eta/s$ never goes below the KSS bound when the baryon chemical potential is finite. Notably, the impact of the finite size of the system on $\eta/s$ fades away at high temperatures for both zero and finite values of baryon chemical potential. 

\par Next, we discuss the bulk viscosity $\zeta_b$, which is an interesting quantity because of its relation to the conformal symmetry of the system \cite{arnold06}. Subplots (c) and (d) of Fig. \ref{vis} depicts the normalised bulk viscosity $\zeta_b/s$ as a function of $T/T_c$.
While $\zeta_b$ is nearly zero in the low-temperature regime, $\zeta_b/s$ starts from high values at low $T$ due to the comparative magnitudes of $\zeta_b$ and $s$. For zero chemical potential, we find that the normalised bulk viscosity $\zeta_b/s$ agrees with lattice data points \cite{meyer2007,meyer2008}. We note that $\zeta_b/s$ gradually decreases to zero at higher temperatures. This is expected as the conformal limit of QCD is expected to be reached at high temperatures and may be attributed to the dominant increase in $s$ compared to $\zeta_b$ at high $T$, indicating that the system becomes conformally symmetric. In subplot (d) of Fig. \ref{vis}, we plot $\zeta_b/s$ for different system sizes $R=\infty$, 5 fm, and 3 fm at baryon chemical potential $\mu_B=600$ MeV, for both Polyakov loops $\cal{U_L}$ and $\cal{U_P}$. We observe that the value of  $\zeta_b/s$ is less compared to the case of zero chemical potential, even at lower temperatures. This may indicate that conformal symmetry is achieved at much lower temperatures for systems with non-zero baryon chemical potential. We should point out here that just like the specific shear viscosity $\eta/s$, the normalised bulk viscosity $\zeta_b/s$ increases ever so slightly with decreasing system size ($R<\infty$), and this effect disappears at high $T$.

    \begin{figure*}
    \centering
    \begin{minipage}[c]{0.98\textwidth}
    (a)\includegraphics[width=7.4cm]{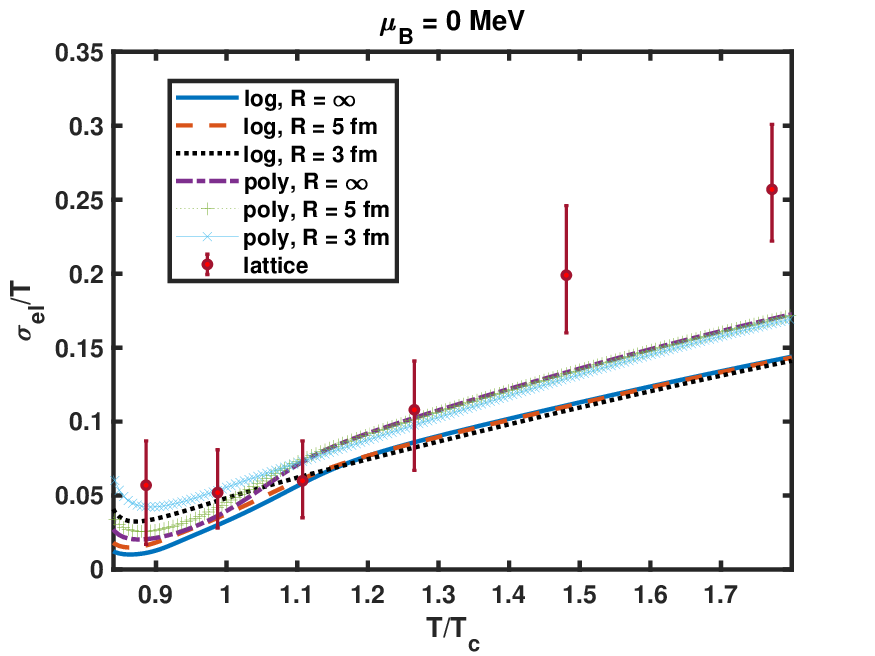}
    \hspace{0.03cm}
    (b)\includegraphics[width=7.4cm]{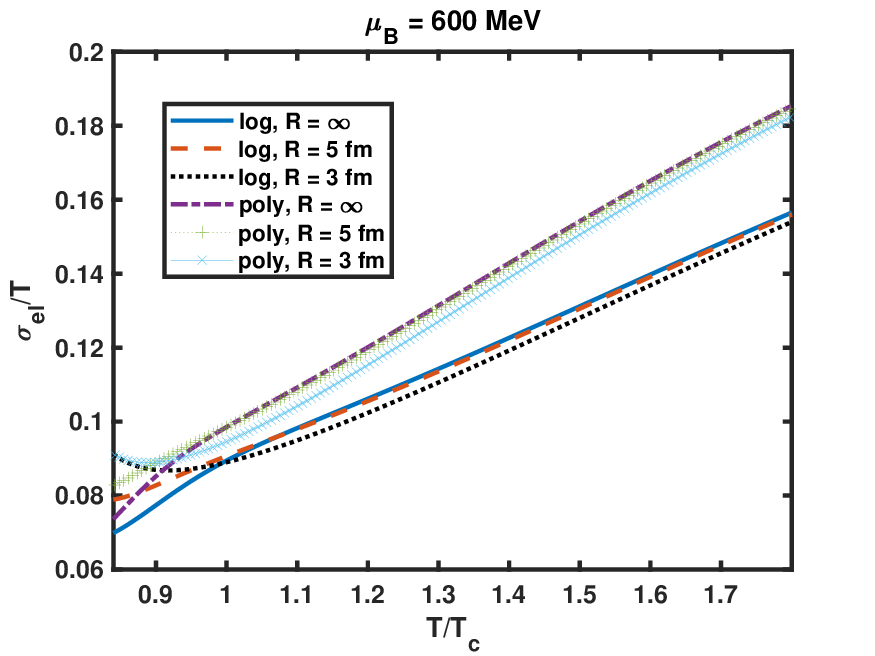}
    \hspace{0.03cm}	
    \end{minipage}
    \caption{\label{sig} (Colour online) The normalised electrical conductivity $\sigma_{el}/T$ as a function of $T/T_c$ for system sizes $R=\infty$, 5 fm, and 3 fm, at $\mu_B = 0$ MeV compared to lattice data from Ref. \cite{Aarts} [in subplot (a)] and $\mu_B = 600$ MeV, $\mu_I = -30$ MeV, and $\mu_S=125$ MeV [in subplot (b)], for Polyakov loop potentials $\cal{U_P}$ and $\cal{U_L}$.}
    \end{figure*}

Next, we study the electrical conductivity $\sigma_{el}$ and the thermal conductivity $\kappa$ in the framework of finite volume PCQMF model. Understanding $\sigma_{el}$ is important as it shows the ease with which electric charge (carried by quarks) can move through the system, making it essential for studying electromagnetic interactions in QCD matter \cite{patra}. Fig. \ref{sig} illustrates $\sigma_{el}/T$ as a function of $T/T_c$ for different system sizes ($R = \infty$, 5 fm, and 3 fm) and two Polyakov loops $\cal{U_L}$ and $\cal{U_P}$ at baryon chemical potentials $\mu_B=0$ and $600$ MeV. The results for zero baryon chemical potential are compared to lattice data points from Ref. \cite{Aarts}. At $\mu_B=0$ MeV, we observe that $\sigma_{el}/T$ starts from a very small value at lower temperatures and rises with temperature. This may be due to the deconfinement of quarks at higher temperatures, enabling them to move freely hence enhancing their conductivity properties. Similar observations for $\sigma_{el}/T$ were reported in the NJL model \cite{ghosh2019} and the PNJL model \cite{saha2018,zhao}. Notably, the value of $\sigma_{el}/T$ matches well with the lattice data at lower temperatures. Regarding the impact of the finite size of the system, we find that $\sigma_{el}/T(R=3$ fm) increases with respect to $\sigma_{el}/T(R=\infty)$ below the transition temperature. 
In Ref. \cite{saha2018}, authors used the PNJL model and found an enhancement in the electric conductivity for smaller system sizes. 
This can be attributed to two factors. One is the decrease in the quark masses: as discussed earlier in Fig. \ref{mass}, decreasing $R$ decreases the effective masses of quarks. Lighter quarks move faster and have a higher $\sigma_{el}/T$ value. Another reason for the enhancement of $\sigma_{el}/T$ for finite size systems may also be the increasing tendency of the quarks to become deconfined at lower temperatures, as pointed out in Fig. \ref{poly} and \ref{polyd}. This unrestricted movement allows the charge carriers to flow more freely, further increasing their value for smaller systems. However, we observe that at temperatures higher than $T_c$, $\sigma_{el}/T$ decreases for smaller systems ($R=3$ fm). For the finite value of baryon chemical potential $\mu_B=600$ MeV in Fig. \ref{sig}(b), we found that $\sigma_{el}/T$ starts from a non-zero value at lower temperatures. Again, this may be due to the shifting of transition and deconfinement temperature to lower values at higher chemical potentials. 
Reducing the system size results in shifting $\sigma_{el}/T$ to higher magnitude for both $\cal{U_L}$ and $\cal{U_P}$, similar to $\mu_B = 0$ MeV case. 
Again, the effect of finite size on $\sigma_{el}/T$ is different before and after the transition temperature, i.e., decreasing the system size leads to an increase in $\sigma_{el}/T$ for temperatures below $T_c$ and a decrease above it.
\begin{figure}
	\centering
	\includegraphics[width=10cm]{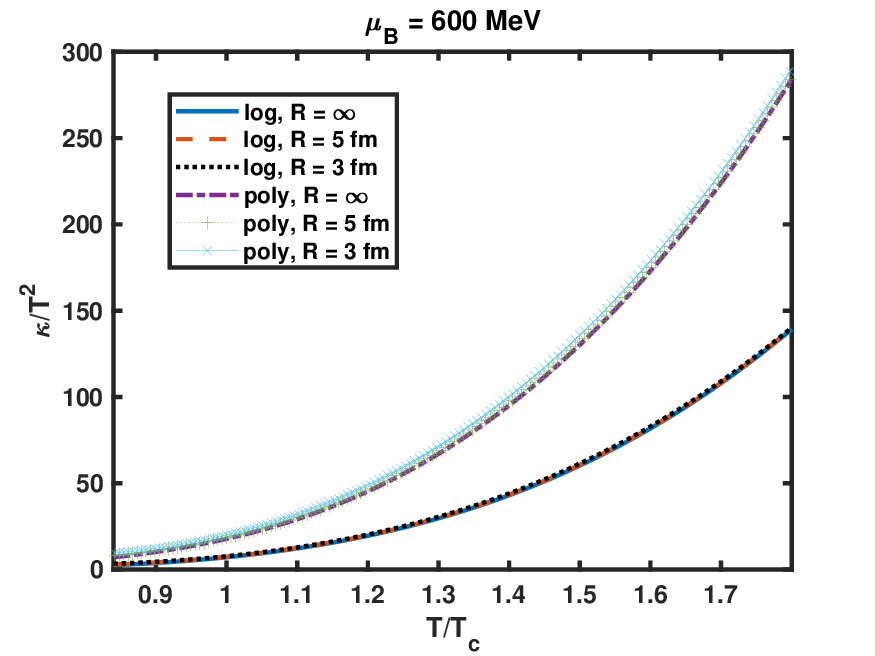}
	\caption{The variation of normalised thermal conductivity $\kappa/T^2$ with $T/T_c$ for system sizes $R=\infty,$ 5 fm, and 3 fm, at $\mu_B = 600$ MeV, $\mu_I = -30$ MeV, and $\mu_S=125$ MeV, for Polyakov loop potentials $\cal{U_P}$ and $\cal{U_L}$.}
	\label{kap}
\end{figure}

\par Another important conductive property is the thermal conductivity $\kappa$, which is related to the heat flow in a relativistic fluid, in our case QGP. It determines the rate of energy change in the system and has received interest among particle physicists \cite{deb,mitra2015,mitra20152}. As was pointed out in Sec. \ref{transport}, $\kappa$ diverges for $\mu_B=0$ MeV due to the diverging nature of the number density $\rho$. Fig. \ref{kap} shows $\kappa/T^2$ as a function of $T/T_c$ for system sizes $R=\infty$, 5 fm, and 3 fm at $\mu_B = 600$ MeV, $\mu_I = -30$ MeV, $\mu_S=125$ MeV, for Polyakov loop potentials $\cal{U_L}$ and $\cal{U_P}$. We find $\kappa/T^2$ to be a monotonically increasing function of $T$. The increase in $\kappa/T^2$ is likely caused by the heat function $h$, which experiences a rise with temperature (not shown in the figure). As the system heats up, the value of $h$ rises, resulting in a smooth transfer of heat within the QGP, thereby boosting thermal conductivity. A similar rise of $\kappa/T^2$ with the temperature has been reported in Refs. \cite{patra,islam,ghosh2019}. Regarding the dependence of $\kappa$ on the system size $R$, we find that just like $\sigma_{el}$, it increases with decreasing value of $R$.


    \begin{figure*}
    \centering
    \begin{minipage}[c]{0.98\textwidth}
    (a)\includegraphics[width=7.4cm]{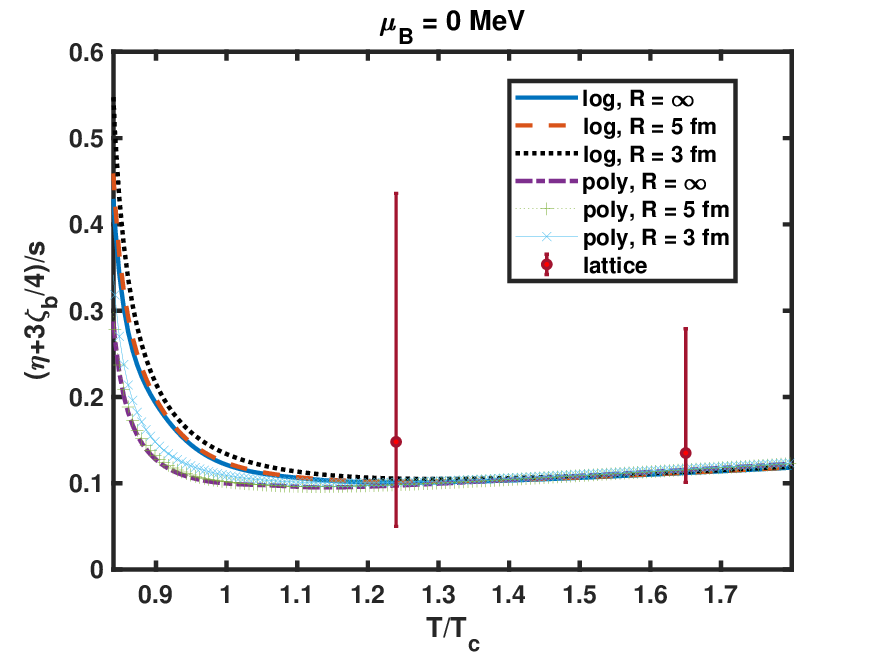}
    \hspace{0.03cm}
    (b)\includegraphics[width=7.4cm]{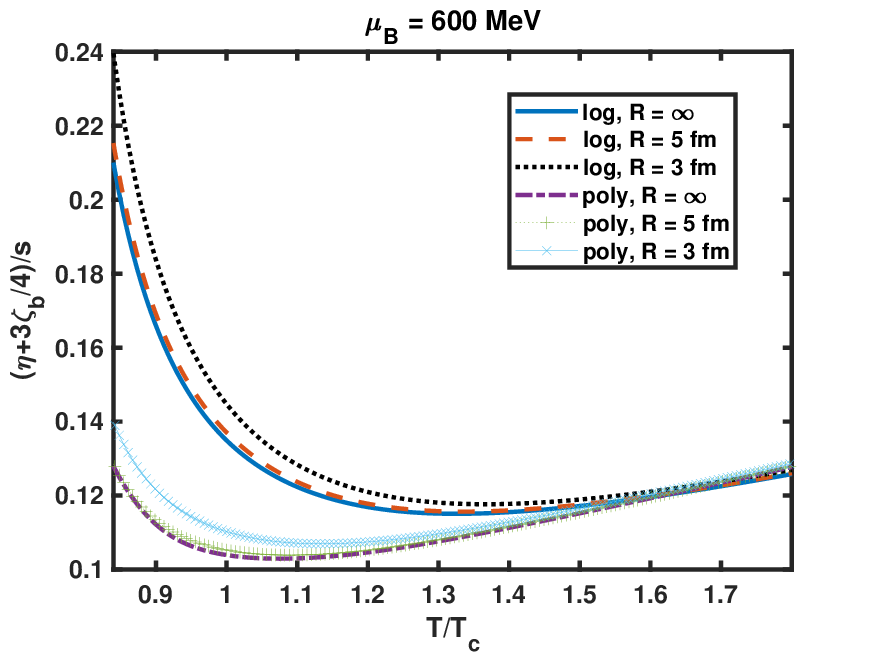}
    \hspace{0.03cm}	
    (c)\includegraphics[width=7.4cm]{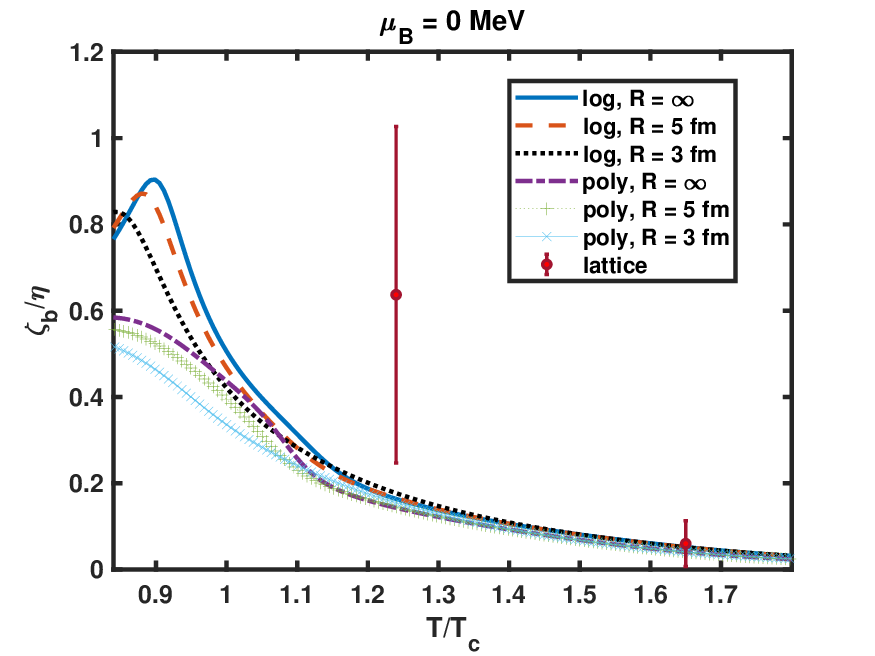}
    \hspace{0.03cm}
    (d)\includegraphics[width=7.4cm]{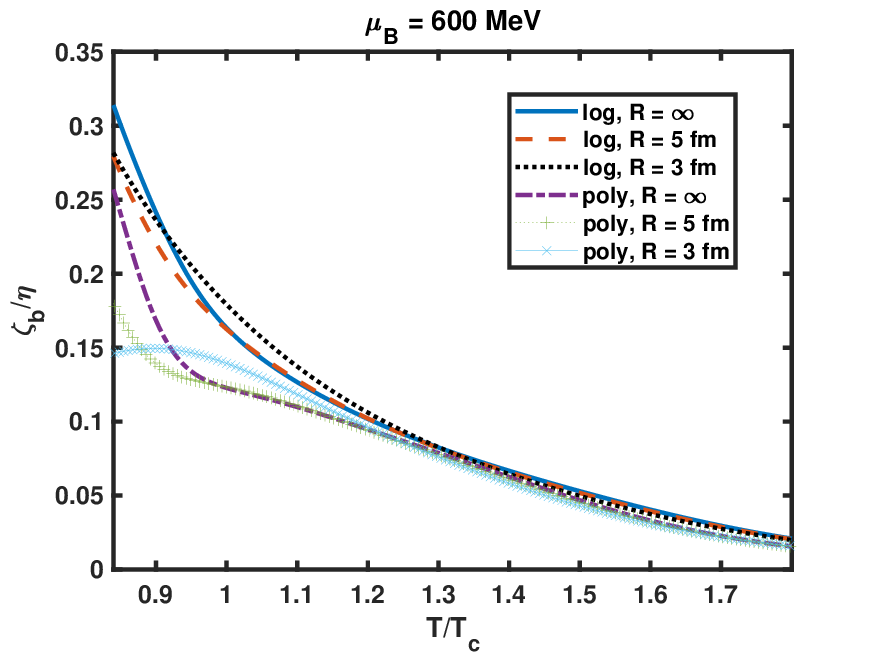}
    \hspace{0.03cm}	
    \end{minipage}
    \caption{\label{visc ratio} (Colour online) The variation of specific sound channel $(\eta+3\zeta_b/4)/s$ and bulk to shear viscosity ratio $\zeta_b/\eta$ with $T/T_c$ are plotted for system sizes $R=\infty$, 5 fm, and 3 fm, at $\mu_B = 0$ MeV compared to lattice data points from Refs. \cite{meyer2007,meyer2008} [in subplots (a) and (c)] and $\mu_B = 600$ MeV, $\mu_I = -30$ MeV, and $\mu_S=125$ MeV [in subplots (b) and (d)], for Polyakov loop potentials $\cal{U_P}$ and $\cal{U_L}$.}
    \end{figure*} 

In Fig. \ref{visc ratio}, we have shown the specific sound channel $(\eta+3\zeta_b/4)/s$ as well as the ratio of bulk to shear viscosity $\zeta_b/\eta$ as a function of $T/T_c$ for varying system sizes ($R=\infty$, 5 fm, and 3 fm) at $\mu_B=0$ and $600$ MeV, for Polyakov loops $\cal{U_L}$ and $\cal{U_P}$. The result for $\mu_B = 0$ MeV is compared with lattice data points from Refs. \cite{meyer2007,meyer2008}. The behaviour of $(\eta+3\zeta_b/4)/s$ is dominated by $\eta/s$, which shows a minimum near the transition temperature for both $\mu_B=0$ and $\mu_B=600$ MeV and then rises slowly at higher temperatures. As discussed earlier, in the high $T$ region, the normalised bulk viscosity $\zeta_b/s$ vanishes; however, the specific shear viscosity $\eta/s$ remains finite. In Ref. \cite{ozven}, the authors reported similar findings using parton-hadron-string dynamics (PHSD) simulations in RTA.
As for the effect of the finite size of the system, we see that for a fixed $T, (\eta+3\zeta_b/4)/s$ is slightly higher for smaller systems. However, we observe that the specific sound channel becomes independent of the system size at higher temperatures. In Fig. \ref{visc ratio}(c), for $\mu_B=0$ MeV, the ratio $\zeta_b/\eta$ shows a peak near the transition temperature, which is more prominent for the Polyakov loop $\cal{U_L}$ as compared to $\cal{U_P}$. We note that the value of $\zeta_b/\eta$ disappears due to the vanishing value of $\zeta_b$ at higher temperatures. Decreasing the size of the system from $R=\infty$ to 3 fm results in shifting the peak towards lower $T$ for both Polyakov loops $\cal{U_L}$ and $\cal{U_P}$. As for the case of finite baryon chemical potential $\mu_B = 600$ MeV in Fig. \ref{visc ratio}(d), we find that the peak fades away. Again, this can be seen as an indication of shifting the transition temperature to lower values at finite chemical potential. 

\section{Summary}
\label{summary}
In this work, we have used the framework of Polyakov chiral SU (3) quark mean field model to study the temperature dependence and the effect of the finite size of the system on various transport coefficients, namely specific shear viscosity $\eta/s$, normalised bulk viscosity $\zeta_b/s$, electrical conductivity $\sigma_{el}/T,$ and thermal conductivity $\kappa/T^2$ at zero 
as well as at finite baryon chemical potential, considering the finite value of isospin and strangeness chemical potentials.
In addition, we have also computed the temperature dependence of various thermodynamic quantities: pressure $p/T^4$, energy density $\epsilon/T^4$, entropy density $s/T^3$, trace anomaly $(\epsilon-3p)/T^4$, speed of sound squared $c_s^2$, and specific heat $c_v/T^3$ and study the impact of finite size on the thermodynamic quantities. All the quantities were studied for the logarithmic ($\cal{U_L}$) and the polynomial form of Polyakov loop potential ($\cal{U_P}$) with quark back reaction at different system sizes $R=\infty$, 5 fm, and $3$ fm. Additionally, the fermionic vacuum term has been incorporated into the PCQMF model. The transport coefficients are obtained using a quasiparticle approach in the kinetic theory under relaxation time approximation (RTA) with temperature and chemical potential-dependent relaxation time. To study the transport coefficients under the finite size consideration, we have employed a lower momentum cutoff in the integral of the thermodynamic potential density of the PCQMF model.
This leads to a reduction in the quark masses and an enhancement in the value of the thermodynamic quantities at temperatures less than the transition temperature $T_c$.

We have also observed that decreasing the system size leads to shifting the pseudo-critical temperature of the chiral transition $T_{\chi}$ and the deconfinement temperature $T_d$ to lower values. Additionally, we found that increasing the baryon chemical potential further reduces $T_{\chi}$ and $T_d$. In addition, we have also studied the dependence of vector coupling $g_v$ on the order of the chiral phase transition and observed that increasing $g_v$ leads to the disappearance of the first order chiral phase transition.
For the transport coefficients, the effect of finite size enters through the integrals, Eqs. (\ref{eta})-(\ref{kappa}). 
The value of specific shear viscosity $\eta/s$ shows a minimum and approaches the KSS bound near the transition temperature and increases slowly afterwards. However, $\zeta_b/s$ vanishes at high $T$.
The electrical conductivity $\sigma_{el}/T$ and the thermal conductivity $\kappa/T^2$ show rises with temperature. Reducing the system size enhances the values of these transport coefficients below $T_c$. We found that the impact of the finite size on the transport coefficients is more prominent in the vicinity of the transition temperature and vanishes for higher temperatures. The specific sound channel $(\eta+3/4\zeta_b)/s$ and the bulk-to-shear viscosity ratio $\zeta_b/\eta$ are also studied. $(\eta+3/4\zeta_b)/s$ shows similar behaviour to $\eta/s$ while $\zeta_b/\eta$ peaks near the transition temperature and vanishes at high $T$. 
We conclude that the derived transport coefficients, shear viscosity ($\eta$), bulk viscosity ($\zeta_b$), electrical conductivity ($\sigma_{el}$), thermal conductivity ($\kappa$) are crucial for the investigation of hot and dense QCD matter. The inclusion of fermionic vacuum term and finite volume in the PCQMF model leads to qualitative and partially quantitative alignment of the quantities, such as subtracted chiral condensate, thermodynamic quantities, as well as transport coefficients with lattice QCD results.

\section{ACKNOWLEDGMENT}
The authors sincerely acknowledge the support toward this work from the Ministry of Science and Human Resources (MHRD), Government of India, via an Institute fellowship under the Dr B R Ambedkar National Institute of Technology Jalandhar.



\end{document}